\documentclass[11pt]{iopart}
\usepackage{amssymb}
\pdfoutput=1
\usepackage{graphicx}

\begin{document}

\title[Catastrophic regime shifts in model ecological communities are true phase
transitions]
{Catastrophic regime shifts in model ecological communities are true phase
transitions}

\author{J.~A.\ Capit\'an\footnote{jcapitan@math.uc3m.es} 
and J.~A. Cuesta\footnote{cuesta@math.uc3m.es}}

\address{Grupo Interdisciplinar de Sistemas Complejos (GISC), 
Departamento de Matem\'aticas, Escuela Polit\'ecnica Superior, 
Universidad Carlos III de Madrid, E-28911 Legan\'es, Madrid, Spain}

\begin{abstract}
Ecosystems often undergo abrupt regime shifts in response to gradual external
changes. These shifts are theoretically understood as a regime switch between 
alternative stable states of the ecosystem dynamical response to smooth changes 
in external conditions. Usual models introduce nonlinearities in the macroscopic 
dynamics of the ecosystem that lead to different stable attractors among which 
the shift takes place. Here we propose an alternative explanation of catastrophic 
regime shifts based on a recent model that pictures ecological communities as 
systems in continuous fluctuation, according to certain transition probabilities,
between different {\em micro-states} in the phase space of viable communities. 
We introduce a spontaneous extinction rate that accounts 
for gradual changes in external conditions, and upon variations on this control parameter
the system undergoes a regime shift with similar features to those previously
reported. Under our microscopic viewpoint we recover the main results obtained 
in previous theoretical and empirical work (anomalous variance, hysteresis cycles, trophic 
cascades). The model predicts a gradual loss of species in trophic levels from bottom 
to top near the transition. But more importantly, the spectral analysis 
of the transition probability matrix allows us to {\em rigorously} establish that 
we are observing the fingerprints, in a finite size system, of a true phase transition 
driven by background extinctions.
\end{abstract}

\submitto{{\it J.~Stat.~Mech.}}
\maketitle

\section{Introduction}\label{s:intro}

Ecosystems are exposed to continuous changes in external conditions. Seasonal 
changes of environmental conditions, climate oscillations, variations in the 
amount of resources and nutrient loading, habitat fragmentation, harvest or loss 
of species diversity are a  few examples of these gradual changes. They often 
change slowly, even linearly, with time \cite{tilman:2001}. It is usually assumed 
that the response of the system to external changes is smooth most of the times. 
However, occasionally sudden changes can occur. For example, the sudden loss of 
transparency and vegetation observed in shallow lakes due to human-induced effects 
\cite{scheffer:2003}; corals overgrown by macro-algae in the Caribbean reef seem 
to shift between two stable states rather than responding smoothly to 
external conditions \cite{done:1991,nystrom:2000}; or in savannahs, sparse trees 
with a grass layer can switch to a dense woody state as a result of the
alternation in fire and grazing regimes \cite{walker:1993,ludwig:1997}.
All these phenomena share the feature that ecosystems seem to change between 
two different stable states. Sudden changes between two regimes are the so-called 
catastrophic shifts \cite{scheffer:2001}. Hence, when subjected to a slowly changing 
external control variable, ecological communities may show little change until a 
critical point is reached. Then a sudden switch to a contrasting state can
occur.

The simplest theoretical explanation to catastrophic regime shifts comes from the
existence of alternative stable states in the dynamical ecosystem response to
gradual changes.  The shift between two alternative stable states is responsible 
for the transition. Often the existence of different stable states
is associated to nonlinearities. A non-linear ecosystem response to smooth external
variations allows for the existence of alternative stable states \cite{scheffer:2001,may:1977}.
The effects of nonlinearities have also been observed in natural communities. For
example, it has been established that the non-linear dynamics of overexploited 
marine ecosystems magnifies the variability in the abundance of exploited 
species \cite{hsieh:2006,anderson:2008}.

In these models, ecosystems are described from a macroscopic viewpoint.
Usually a global magnitude, representative of the whole community, is used as 
fundamental variable (for instance, the total biomass density). Models are 
basically devised to describe the time evolution of this magnitude by means of 
non-linear functional responses \cite{scheffer:2001,ludwig:1978}. Recently this
paradigm has been applied to spatially extended interacting communities 
\cite{fernandez:2009,dakos:2010}. This theoretical approach is conceptually very similar
to the traditional thermodynamic explanation of phase transitions in physical 
systems, such as the liquid-vapor transition. The lack of convexity of 
theoretical thermodynamic potentials ---such as free energy--- leads to alternative
stable states corresponding to liquid and vapor and both phases coexist
below a certain critical temperature. In 
\cite{ludwig:1978,fernandez:2009,dakos:2010} the analogy is
very clear. The dynamics of biomass density follows a logistic growth with
carrying capacity $K$ and a density-dependent consumption term modelled as
a sigmoidal (Holling's type III) functional response. It is precisely this 
type of functional response which allows for the existence of two separate,
stable equilibrium points in the dynamics above a certain critical value of the
carrying capacity.

However, the current understanding of phase transitions in physical systems 
goes beyond phenomenological, macroscopic models. The microscopic approach 
of Statistical Mechanics represents a more fundamental way of understanding
phase transitions, and many elaborated theories have been developed to
account for abrupt shifts in physical systems. Our aim in this paper is
to propose an alternative explanation, based mainly in a microscopic
approach, to these catastrophic regime shifts in ecosystems. In fact, we will
show that a linear functional response in the dynamics of individual species
can lead to a global shift in the ecosystem between a species-rich attractor
and a state with low species richness. The early-warnings that
are usually mentioned as precursors of catastrophic regimes, like the increasing
fluctuations near the transition \cite{hsieh:2006,scheffer:2009} or the appearance
of trophic cascades \cite{daskalov:2007}, will be recovered within our framework.

Based on an assembly model introduced recently \cite{capitan:2009,
capitan:2010a,capitan:2010b}, that pictures ecosystems as complex, stable
entities which keep on fluctuating between different {\em micro-states}
through successional invasions of rare species, we will introduce a background 
species extinction rate accounting for the gradual, external variations to 
which natural communities are subjected. We will find a threshold rate above which 
the system undergoes a phase transition, and the rate will play the role of the shift
control parameter. We will show that fluctuations become
critical in the vicinity of the transition, and that the ultimate
collapse of the ecosystem correspond to a gradual loss of species from bottom
to top. It is worth remarking that trophic cascades have been recognized 
as signals of overexploitation in marine communities \cite{daskalov:2007}.

The main feature of the assembly model presented in \cite{capitan:2009} is to
provide a complete characterization of the phase space of the system. The time evolution 
of the system is fully described by means of a Markov chain. Under the effect
of increased spontaneous extinctions, an initially species-rich ecosystem can move 
to a region of the phase space where the species richness decreases. Through a
spectral analysis of the transition probability matrix we will be able to
show {\em rigorously} that the shift we observe corresponds to the ``trace''
of a true phase transition ---according to the definition
of Statistical Mechanics--- in finite size.

The paper is organized as follows. In Section \ref{s:extinctions} we will review
the main features of the assembly model and discuss the effect of a background
species extinction rate in the ecosystem. We will find that the Markov process
is ergodic and that there is an increasing probability of finding the ecosystem
close to extinction as the rate grows. In Section \ref{s:signals} we will
study some signals with anomalous behavior near the shift, and show under which
conditions hysteresis in the average number of species arise. In Section
\ref{s:phase} we will show that the observed behavior would correspond to a
true phase transition were the system infinitely large. So we can conclude
that what we are actually observing are the fingerprints of this transition
in systems of finite size.

\section{Background species extinction}\label{s:extinctions}

Our starting point is a simple assembly model that we have recently introduced 
\cite{capitan:2009}. Let us briefly summarize its main features. We describe
ecological communities introducing three basic simplifications. First, species
are arranged in a finite number of trophic levels, and feeding relationships
take place exclusively between contiguous trophic levels. Second, the strength 
of interactions is averaged over the whole community (mean-field), thus
all species at the
same level are trophically equivalent. And third, population dynamics
is modelled by Lotka-Volterra equations. In spite of these oversimplifications,
the resulting communities reproduce all results previously found in
earlier assembly models \cite{capitan:2009,law:1996,morton:1997}.

The mean-field assumption allows us to represent each community by the set of
occupancy numbers of each trophic level, $\{s_{\ell}\}_{\ell=1}^L$, $L$ being 
the total number of levels and $s_{\ell}$ the number of species at level $\ell$.
All species are considered as consumers, with
uniform intrinsic average mortality rate $\alpha$. Predation between adjacent levels 
is modelled by constants $\gamma_+$ and $\gamma_-$. The former represents the 
per-capita rate of increase in population due to feeding, the latter accounts for 
the mean damage caused by being predated. Communities are sustained by a 
primary, abiotic resource characterized by its intrinsic growth rate $R$. This 
represents the saturation value that the resource reaches at equilibrium in absence 
of predation, and measures the energy influx available for each community. 
Species at the first trophic level predate on this resource and transfer the
energy upwards in the food web. Increasing $R$ has the effect of allowing more 
trophic levels and more species in each level as well (see \cite{capitan:2009}). 

Population dynamics is ruled by Lotka-Volterra equations, which guarantees the
existence of a unique interior equilibrium point. Equilibrium communities whose 
populations are above of a given extinction threshold $n_c$ are considered as 
viable. For a full account on the population dynamics and the implications of 
the invasion dynamics in this model, we refer the reader to \cite{capitan:2010a}.

Our model represents a substantial improvement with respect to former assembly 
models in that it provides a complete characterization of all possible invasion 
pathways within the set of viable communities. Equivalently, an
assembly graph $\mathcal G$, i.e. the graph whose nodes are viable communities and 
whose links connect communities through invasions, can be fully
obtained under these assumptions. The existence of a extinction threshold
renders this graph finite. We assigned to each link a certain transition
probability dependent on an invasion rate $\xi$. The mean time between
consecutive invasions, $\xi^{-1}$, is assumed to be large compared to the
time that each community needs to restore its dynamical equilibrium. 

Weighting the links of $\mathcal G$ with 
probabilities defines a finite Markov chain over the set of viable communities. 
In \cite{capitan:2010b} we showed that this chain is aperiodic, hence communities 
can be either transient or recurrent. We always found a single connected set of 
recurrent communities upon increasing the resource saturation $R$. This set was 
either a single absorbing community or a more complex end state of recurrent 
communities, the Markov process being 
ergodic over those complex sets. The existence of complex end states is 
probably one of the most remarkable results of the model (for a detailed 
account of results see \cite{capitan:2010b}).

According to this picture, ecosystems evolve though successional invasions until 
reaching a final end state, either a single absorbent community or a complex, 
closed set of recurrent communities. When the process reaches a complex end state, 
successional invasions transform the ecosystem into some other of the
communities belonging to that set, so the process visits all communities in
this set, albeit with different frequencies 
---given by the asymptotic probability distribution of the Markov process 
within this set. In terms of species this means that communities keep
continuously changing and eventually, after enough
time has elapsed, all the original species in the ecosystem will be replaced by new 
ones. Therefore the ecosystem keeps on fluctuating between the different communities 
comprising the closed set, which persists as a robust, stable entity.

Let us now introduce a rate of spontaneous extinctions. Species in natural 
communities are often subject to overexploitation. Intensive hunting in terrestrial 
communities or the increasing fishing pressure in marine ecosystems are good examples 
of this. Sometimes the species population is seriously altered due to habitat
destruction, in other cases due to exposure to epidemics or diseases. Many
effects like these ones can effectively decrease to critical levels the number
of individuals of a certain species 
or even cause its extinction. We will represent these situations by 
means of a probability rate, $\eta$, which accounts for the probability per unit time 
for a species to go extinct for reasons other than being eaten. Actually 
this probability rate should depend on the species and its environment but,
for the sake of simplicity, we will assume it uniform for all species.

Our model is amenable to introduce background extinctions  in a simple way. 
Elementary processes in the original model for $\eta=0$ 
were transitions between viable communities carried out by single-species invasions. 
Now two different processes, either invasion or extinction, can connect two communities. 
Thus, for a given community $E$ with $L$ trophic levels, we need to determine all the 
possible transitions carried out by invasions and spontaneous extinctions at each level 
(and by invasions at level $L+1$ as well). The graph $\mathcal{G}$
will now contain both types of transitions.
The links corresponding to extinction transitions can be obtained just as the
invasion ones \cite{capitan:2009}. Given a community $E\in\mathcal{G}$, we randomly 
remove one of its species and calculate the equilibrium population densities of the 
resulting community $E'$. If the community is viable, then we establish a transition 
between $E$ and $E'$. If some species go below the extinction level $n_c$, then we 
apply the same sequential extinction procedure that was followed to obtain the
invasion graph (for a detailed discussion of this procedure see \cite{capitan:2010a}). 
We repeat these sequential extinctions until the final community $E''$ is viable.

\begin{figure}[t!]
\begin{center}
\raisebox{70mm}{(a)}\includegraphics[width=48mm,clip=true]{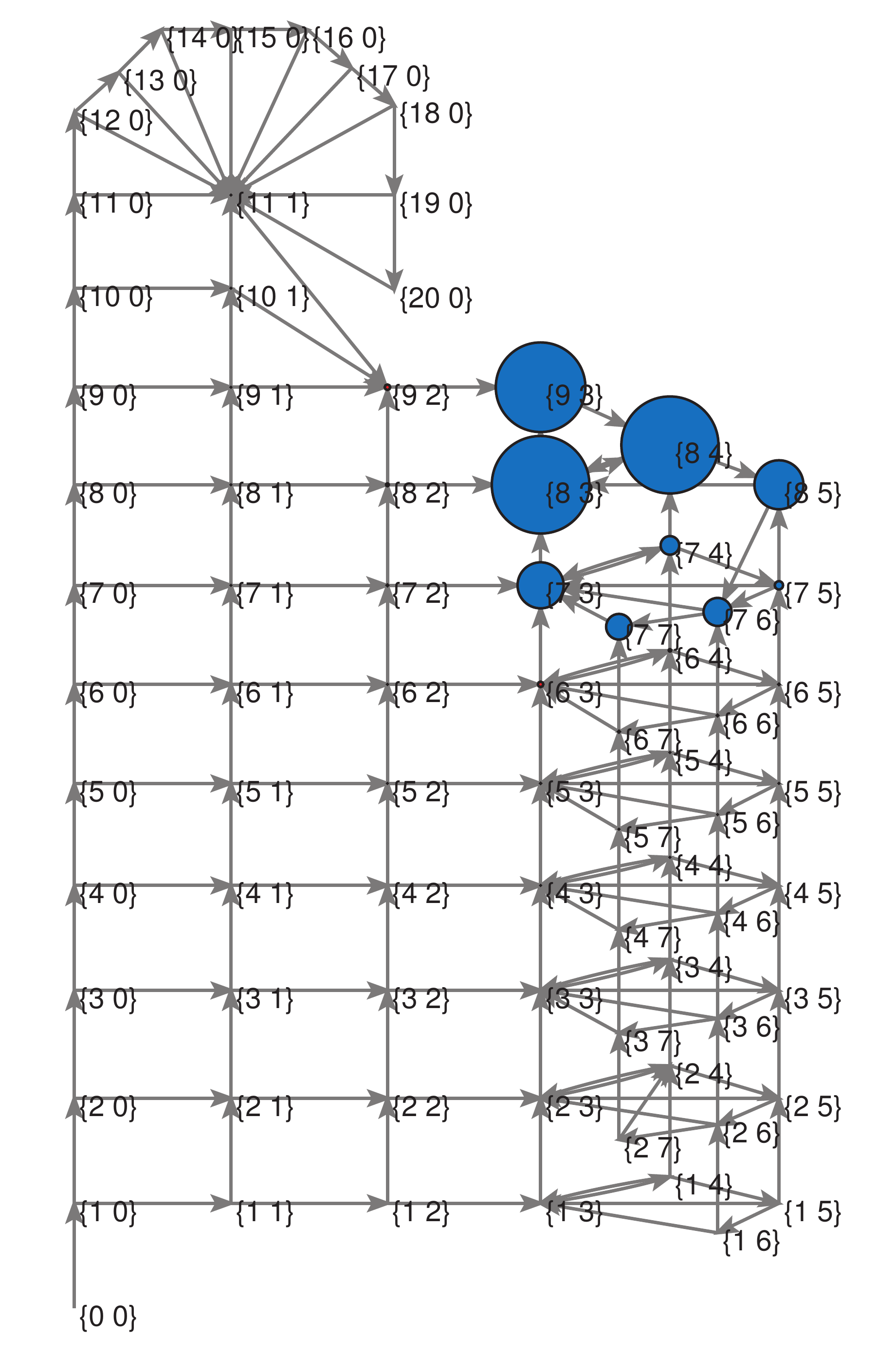}\hspace*{15mm}
\raisebox{70mm}{(b)}\includegraphics[width=48mm,clip=true]{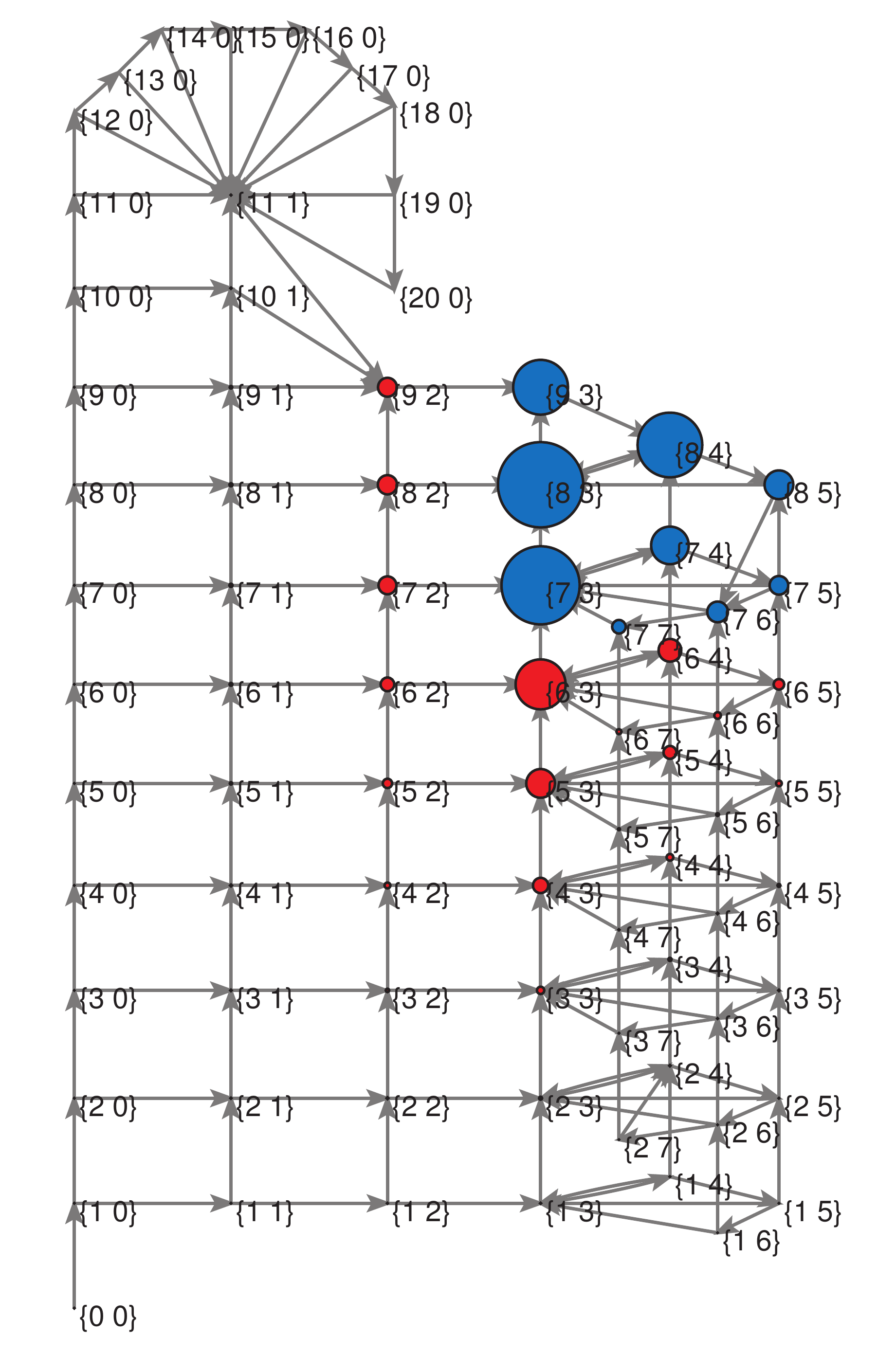}\\
\raisebox{70mm}{(c)}\includegraphics[width=48mm,clip=true]{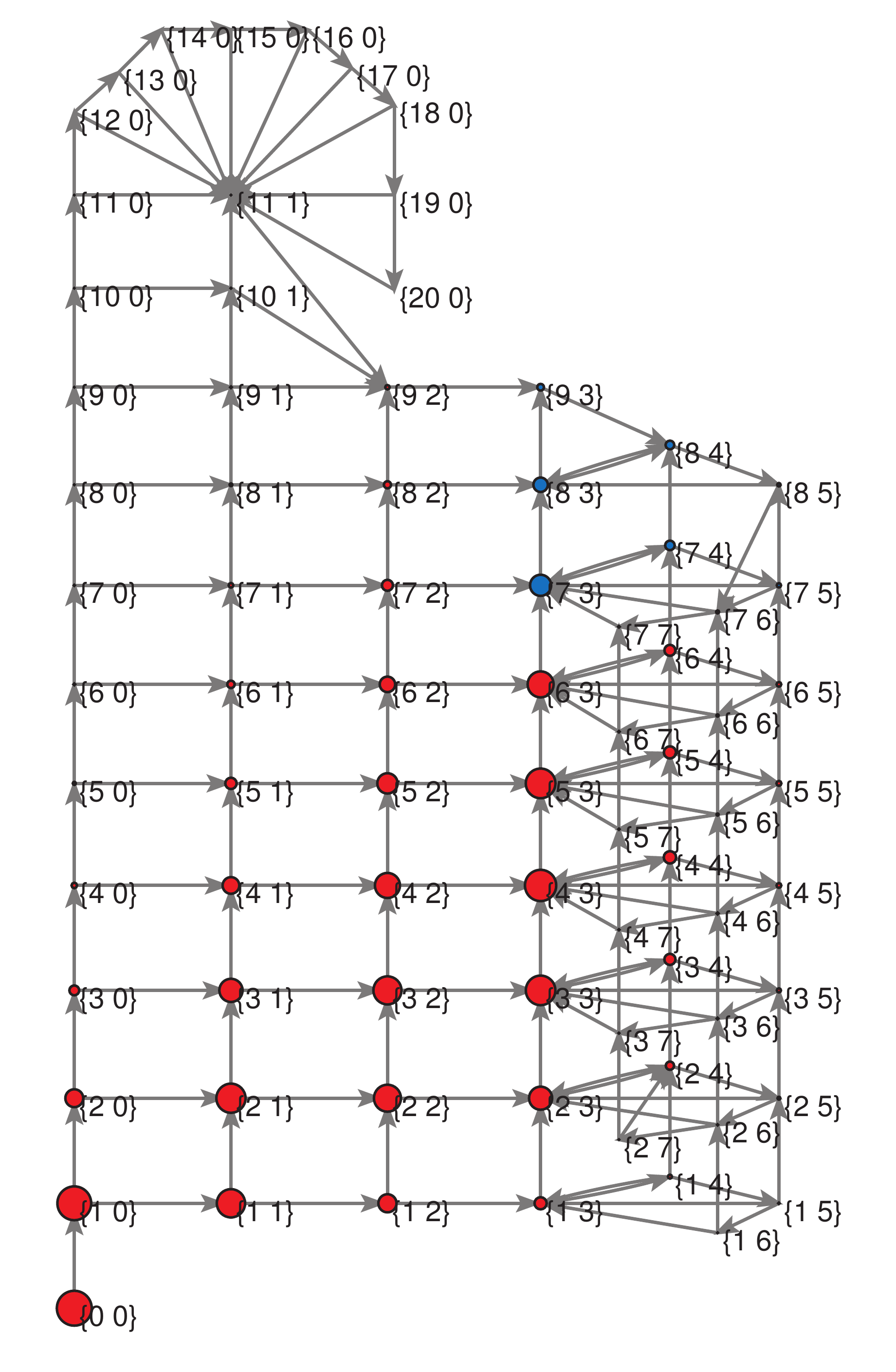}\hspace*{15mm}
\raisebox{70mm}{(d)}\includegraphics[width=48mm,clip=true]{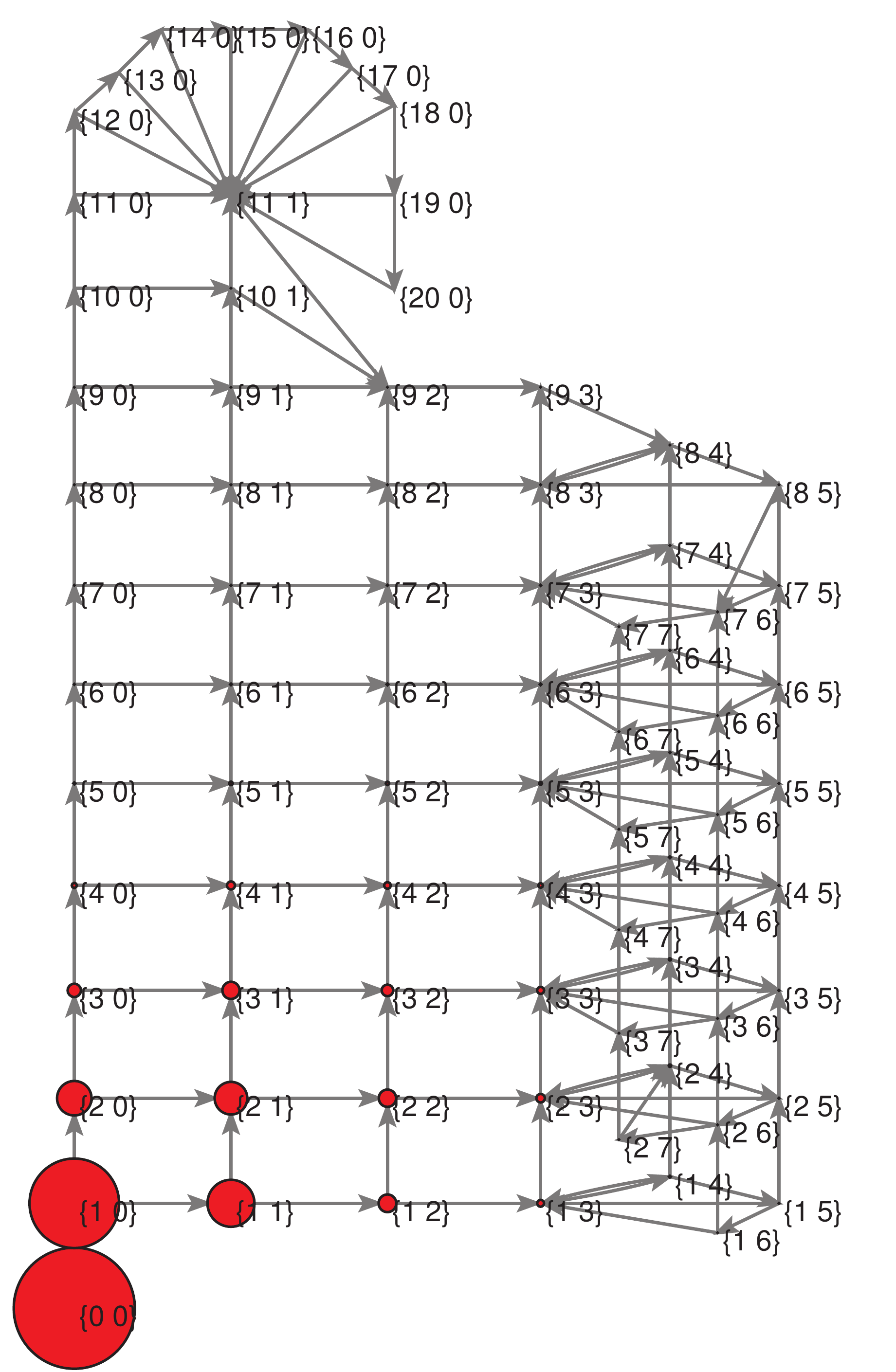}
\caption{Assembly graph $\mathcal{G}$ obtained for $R=120$,
comprising 79 communities with up to 2 trophic levels, for 4 increasing values
of $\eta/\xi$. Diameter of nodes is proportional to their asymptotic probability.
Labels of nodes represent species occupancies $\{s_1,s_2\}$. For the sake of
clarity, only transitions carried out through invasions are shown. For $\eta=0$,
the recurrent set contains 9 nodes (colored in blue). 
Red nodes represent transient states for $\eta=0$.
(a) The most probable
communities are those in the recurrent set ($\eta/\xi=0.05$). (b) Some
communities, close to this set, are visited with high frequency ($\eta/\xi=0.3$).
(c) Almost none of the 9 originally recurrent communities are visited ($\eta/\xi=
0.6$). (d) The most probable community corresponds to the total extinction state 
($\eta/\xi=1$).}
\label{fig:graphs}
\end{center}
\end{figure}

The transition probability $p_{ij}$ for the transition from community $i$
to community $j$ can be written as
\begin{equation}\label{eq:mat}
p_{ij} = \delta_{ij} + \xi q_{ij} + \eta u_{ij},
\end{equation}
where $\delta_{ij}$ is the Kronecker delta, and matrices $Q=(q_{ij})$ and $U=(u_{ij})$ 
account for the relative frequency of invasions and extinctions, respectively.
For $i\ne j$ we set $q_{ij} = n_{ij}/(L+1)$,
$n_{ij}$ being the number of different invasions of $i$ that lead to $j$
and $L+1$ the total number of possible invasions of $i$, provided it has $L$ trophic
levels. For $i\ne j$ we define
$u_{ij} = m_{ij}/S_i$, where $m_{ij}$ is the number of different extinctions in $i$ that 
lead to $j$ and $S_i=\sum_{\ell=1}^L s_{\ell}^{(i)}$ 
is the number of possible extinctions of $i$. We set $u_{ii}=0$ 
and calculate the diagonal of $Q$ so that $P=(p_{ij})$ is a stochastic matrix.
This yields
\begin{equation}
q_{ii} = -\sum_{i\ne j}\left(q_{ij}+\frac{\eta}{\xi}u_{ij}\right).
\end{equation}
When $\eta=0$ we recover the original transition matrix of our model (see
\cite{capitan:2010b}). This is quite a singular case, though, not
representative of what happens for any $\eta>0$ ---no matter how small.

In fact, there is a major difference between the cases $\eta=0$ and $\eta>0$,
regarding the properties of the Markov chain. For any $\eta > 0$, there is a
{\em non-zero} probability for all the $S$ 
species in a community to go extinct. Let $\Delta t$ be the time unit between
consecutive iterations of the Markov chain. Thus the removal of all species caused 
by sequential spontaneous extinctions has a probability at least equal to $(\eta 
\Delta t)^S$. The 
non-vanishing probability of total extinction implies that the process can return to 
the initial state (the empty community $\varnothing$) and therefore the Markov
chain becomes
ergodic. This has to be compared with the former model ($\eta = 0$), for which we 
just found a tiny fraction of recurrent states, and almost all the communities in the 
assembly graph were transient \cite{capitan:2009,capitan:2010b}.

\begin{figure}[t!]
\begin{center}
\includegraphics[width=80mm,clip=true]{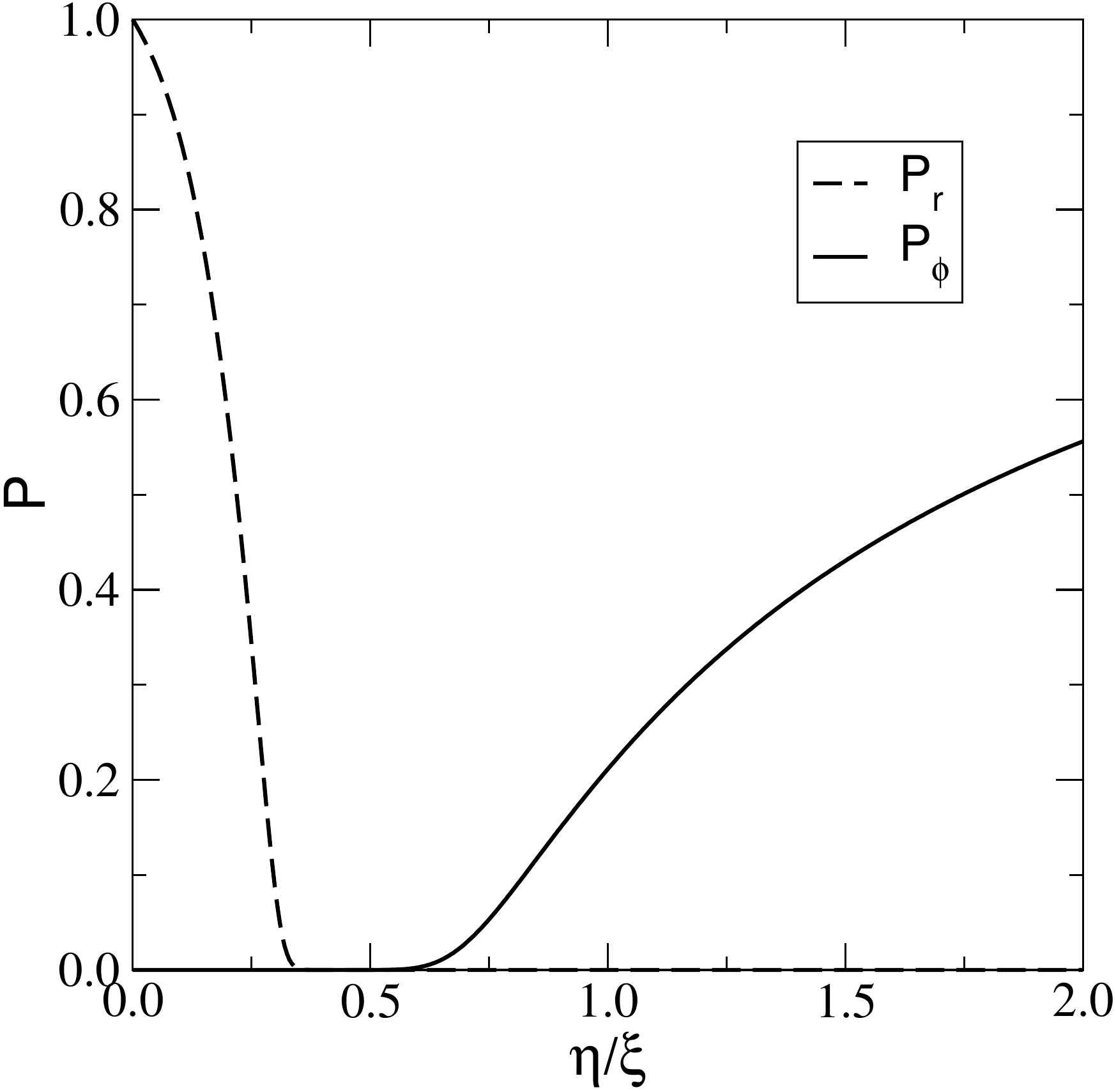}
\caption{Probability $P_{\rm r}$ of finding the process in one of the communities of 
the recurrent set for $\eta=0$ (dashed line), and probability $P_{\varnothing}$ of the 
empty community (full line), as functions of the ratio $\eta/\xi$, for $R=1340$.}
\label{fig:prob}
\end{center}
\end{figure}

Ergodicity implies that any possible state of the ecosystem can be reached with a 
non-zero ---albeit sometimes small--- asymptotic probability. 
According to \eref{eq:mat} we simply need to solve the linear system
\begin{equation}
0=\pi\left(Q+\frac{\eta}{\xi} U\right)
\end{equation}
to obtain the (row) vector $\pi$ of asymptotic probabilities $\pi_i$ for all 
$i\in\mathcal{G}$. Therefore the asymptotic distribution 
depends on the relative strength between rates. We expect 
that, when this ratio is small enough, the subset of communities with the highest 
probability coincides with the recurrent subset found for $\eta=0$. However, as 
this ratio increases, the probability of finding the process within this subset 
should decrease. This effect can be observed in \Fref{fig:graphs}, where we plot 
$\mathcal{G}$ with its asymptotic distribution for $R=120$ and 4 values of the quotient 
$\eta/\xi$. The remaining parameters of the model have been set as in our previous work: 
$\alpha=1$, $\gamma_+=0.5$, $\gamma_-=5$, $\rho=0.3$ and $n_c=1$. (We will use this set 
of parameters throughout this paper.) As $\eta/\xi$ increases, communities that
were recurrent for $\eta=0$ are visited with decreasing asymptotic 
probabilities. Eventually, when the ratio is large enough, these communities are 
hardly visited and the process stays with high probability in communities close to 
the empty ecosystem, $\varnothing$.

This effect is more clearly seen in terms of the dependence of $P_{\rm r}$
---the probability of finding the process in any of the communities of the
recurrent set for $\eta=0$--- and $P_{\varnothing}$ ---the probability of
finding the ecosystem extinct--- on $\eta/\xi$. A typical behavior of these
probabilities is depicted in \Fref{fig:prob}. This plot corresponds to a 
resource saturation $R=1340$, for which $\mathcal{G}$ has 397698 nodes and 539
recurrent 
communities. In \Fref{fig:prob} we can observe an abrupt decrease of $P_{\rm r}$ 
at $\eta/\xi\approx 0.33$, and $P_{\varnothing}$ increases abruptly as well when 
$\eta/\xi\approx 0.65$. Needless to say, these two magnitudes resemble the typical
behavior of order parameters in the vicinity of a phase transition.
A small increase in $\eta$ causes a shift from the stable, recurrent set at 
$\eta=0$ to communities close to extinction. In this sense, increasing background 
extinctions drive the system from a stable, species-rich attractor to a species-poor
region of the phase space. The system thus undergoes a catastrophic regime shift 
analogous to those commonly observed in overexploited ecological communities 
\cite{scheffer:2003}--\cite{ludwig:1997}. 

\section{Signals of catastrophic regime shifts}\label{s:signals}

\begin{figure}[t!]
\begin{center}
\includegraphics[width=88mm,clip=true]{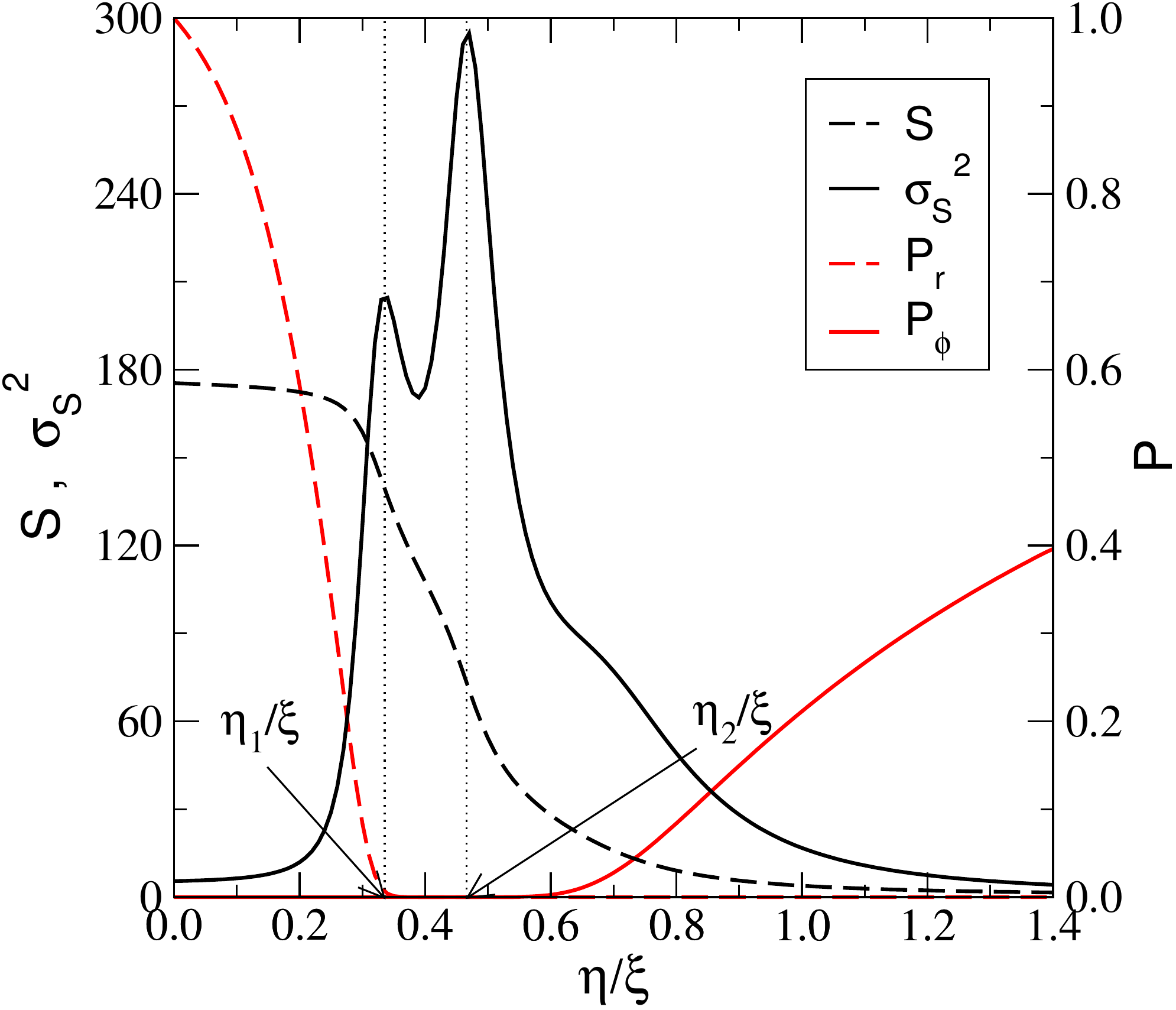}
\caption{Average number of species (black dashed line) and its
variance (black full line) as functions of $\eta/\xi$, for $R=1340$. The
average number of species decreases monotonically whereas its variance
exhibits a double peak at values $\eta_1/\xi\approx 0.33$ 
and $\eta_2/\xi\approx 0.46$. The first one coincides with the abrupt drop of 
$P_{\rm r}$ (red dashed line) but the second one precedes the increase of
$P_{\varnothing}$ (red full line) announcing it.}
\label{fig:fluct}
\end{center}
\end{figure}

From the perspective of conservation and management of ecosystems, it is very important 
to determine some signals that may alert of the proximity of a catastrophic 
transition. These are the so-called early-warnings of catastrophic regime shifts 
\cite{scheffer:2009}, and act as flags for the approach of a critical 
threshold. Although our model is minimalistic, the phenomenology of several magnitudes 
reveal a critical behavior near the shift described in the previous section. 
Now, upon gradually increasing the external ``stress'' (i.e., the ratio
$\eta/\xi$) on our system, 
we will observe abrupt changes in these magnitudes close to the regime shift.

We shall begin assuming very slow variations in $\eta/\xi$, i.e., the ecosystem
undergoes very many invasions before changes in the control parameter are
noticeable. In this situation we can assume that the community is always at is
steady state. At the end of this section we will analyze the effect of relaxing
this assumption and allowing for a mixing of these two time-scales: the scale
of variation of the stress and the scale of invasion.

A first precursor of the shift is the fluctuation of the mean 
number of species in the ecosystem. 
In \cite{fernandez:2009} fluctuations were measured by the 
spatial heterogeneity of a single magnitude (a biomass density)
representing the community as a whole, whereas the kind of fluctuations we are 
considering here are due to changes in the average number of species 
induced by invasions and spontaneous extinctions.
In \Fref{fig:fluct} we plot the average number 
of species $S=\sum_{i\in\mathcal{G}}\pi_iS_i$, where $S_i$ is the total species
richness of the $i$-th community. Its fluctuations are measured by the variance 
\begin{equation}
\sigma_S^2 =\sum_{i\in\mathcal{G}}\pi_iS_i^2-S^2.
\end{equation}
The rapid growth of fluctuations provides an alert of the proximity of the
catastrophic shift \cite{scheffer:2001,fernandez:2009,scheffer:2009} 
Fluctuations for $R=1340$ exhibit 
a double peak at $\eta_1\approx 0.33\xi$ and $\eta_2\approx 0.46\xi$. The first one 
is related to the abrupt drop of the probability $P_{\rm r}$ that we showed in Section 
\ref{s:extinctions}. In addition, these two peaks correspond to a gradual decrease 
in the number of species at each trophic level, as we will show below.

\begin{figure}[t!]
\begin{center}
\includegraphics[width=88mm,clip=true]{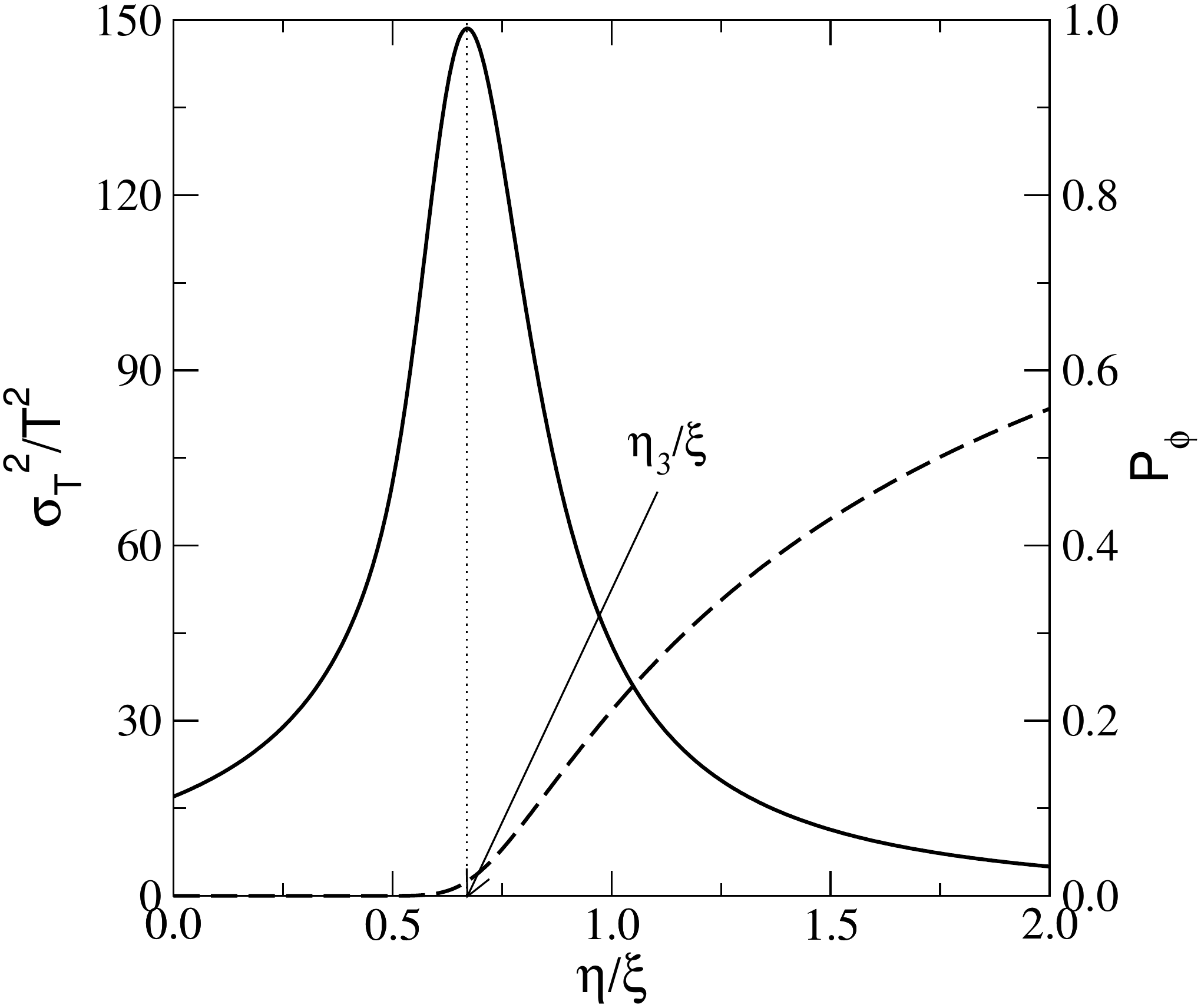}
\caption{Relative variance $\sigma_T^2/T^2$ of the first-return time (average
$T=P_{\varnothing}^{-1}$) to the empty ecosystem (full line), compared to the
probability of total extinction $P_{\varnothing}$ (dashed line), for $R=1340$.
The maximum of the relative variance roughly coincides with the point at which
$P_{\varnothing}$ starts to increase. This maximum 
is reached at $\eta_3/\xi\approx 0.67$.}
\label{fig:fluctTret}
\end{center}
\end{figure}

The second peak at $\eta_2/\xi$ announces the abrupt increase of $P_{\varnothing}$,
but does not coincides with it. The increase of this probability is connected
to the fluctuations of the time of first return to the empty community,
whose mean value is given by $T=P_{\varnothing}^{-1}$. 
Using the first-passage distribution of the Markov chain \cite{feller:1968}, we can
calculate the relative variance $\sigma_T^2/T^2$ (see Appendix for details) and the
result is shown in \Fref{fig:fluctTret}. The maximum relative fluctuation occurs
nearly at
the onset of increase of $P_{\varnothing}$, $\eta_3\approx0.67\xi$. We thus expect 
that relative fluctuations in the average return time to 
any state $i$ close to $\varnothing$ will be amplified close to the extinction 
transition. This notwithstanding, it is hard to figure out how this
fluctuation could be used in practice as a signal of the catastrophe.

\begin{figure}[t!]
\begin{center}
\raisebox{36mm}{(a)}\includegraphics[width=62mm,clip=true]{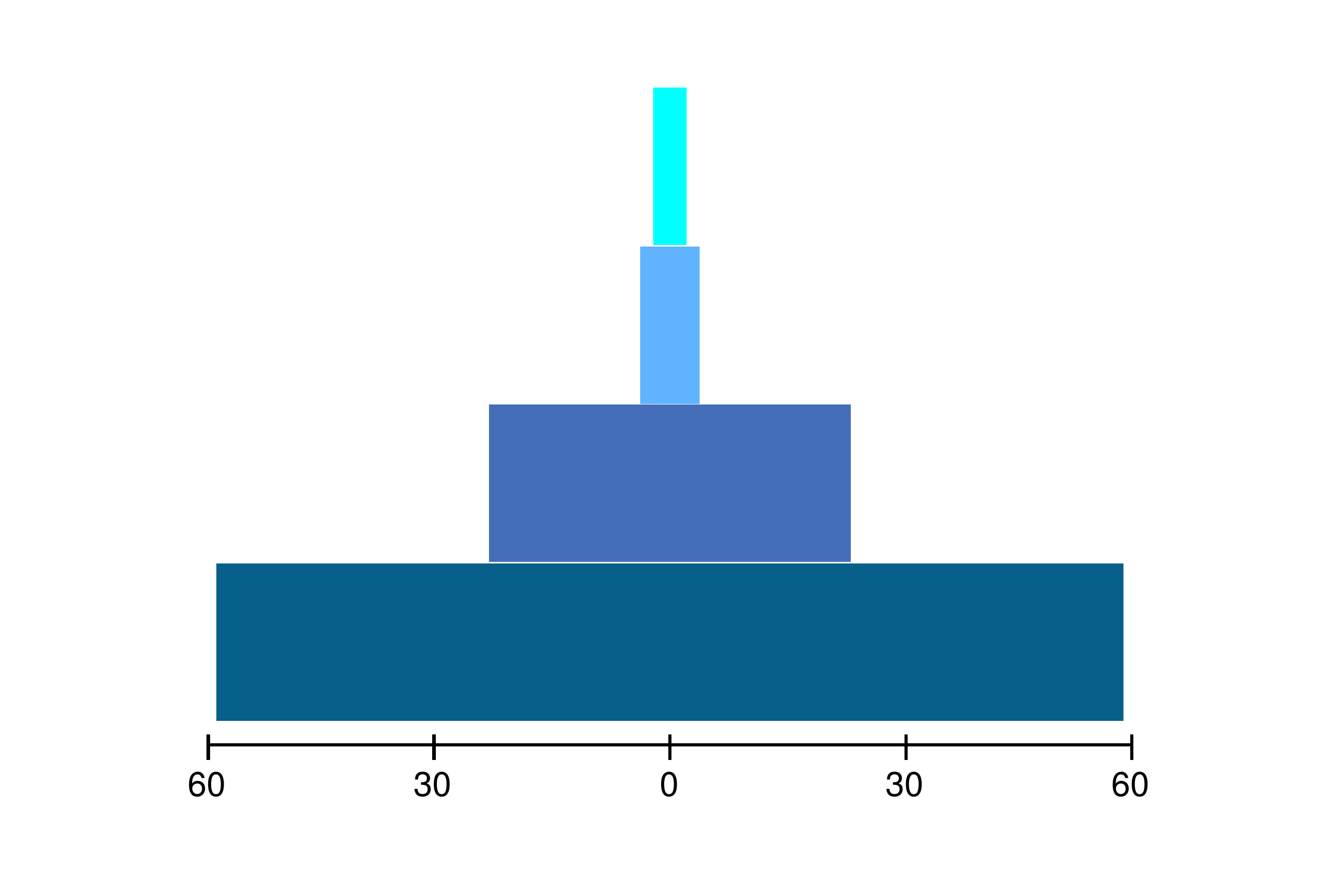}\hspace*{4mm}
\raisebox{36mm}{(b)}\includegraphics[width=62mm,clip=true]{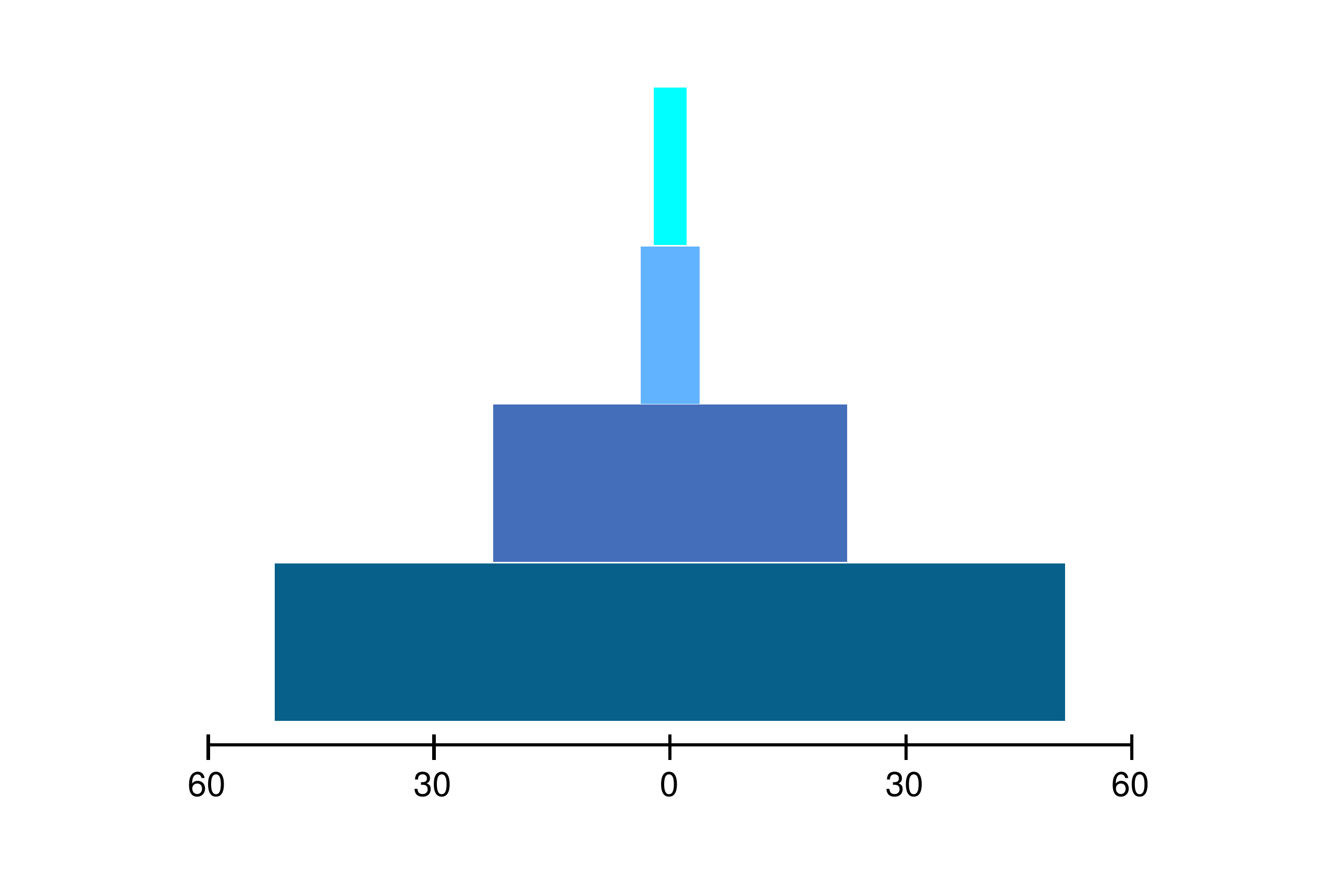}\\
\raisebox{36mm}{(c)}\includegraphics[width=62mm,clip=true]{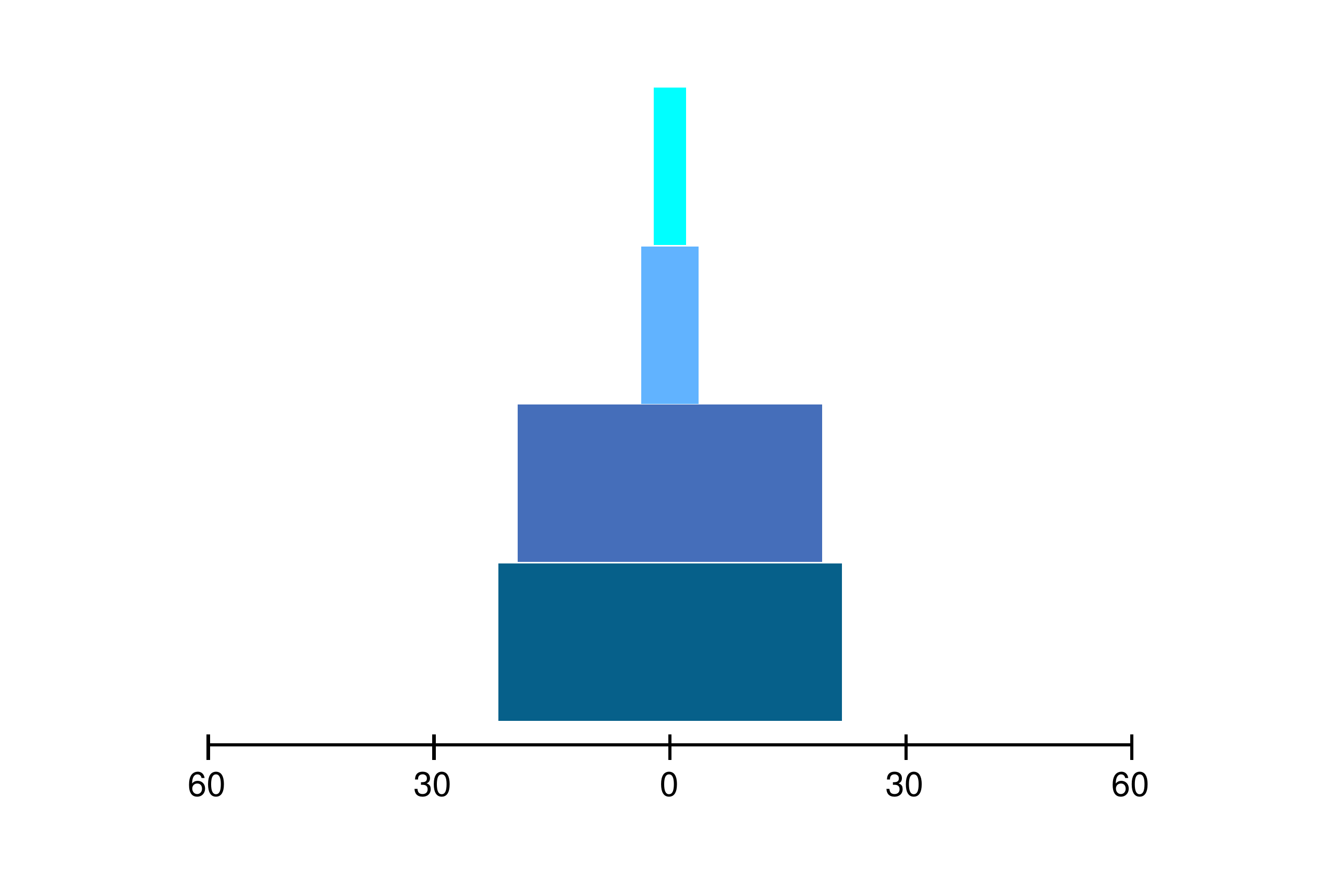}\hspace*{4mm}
\raisebox{36mm}{(d)}\includegraphics[width=62mm,clip=true]{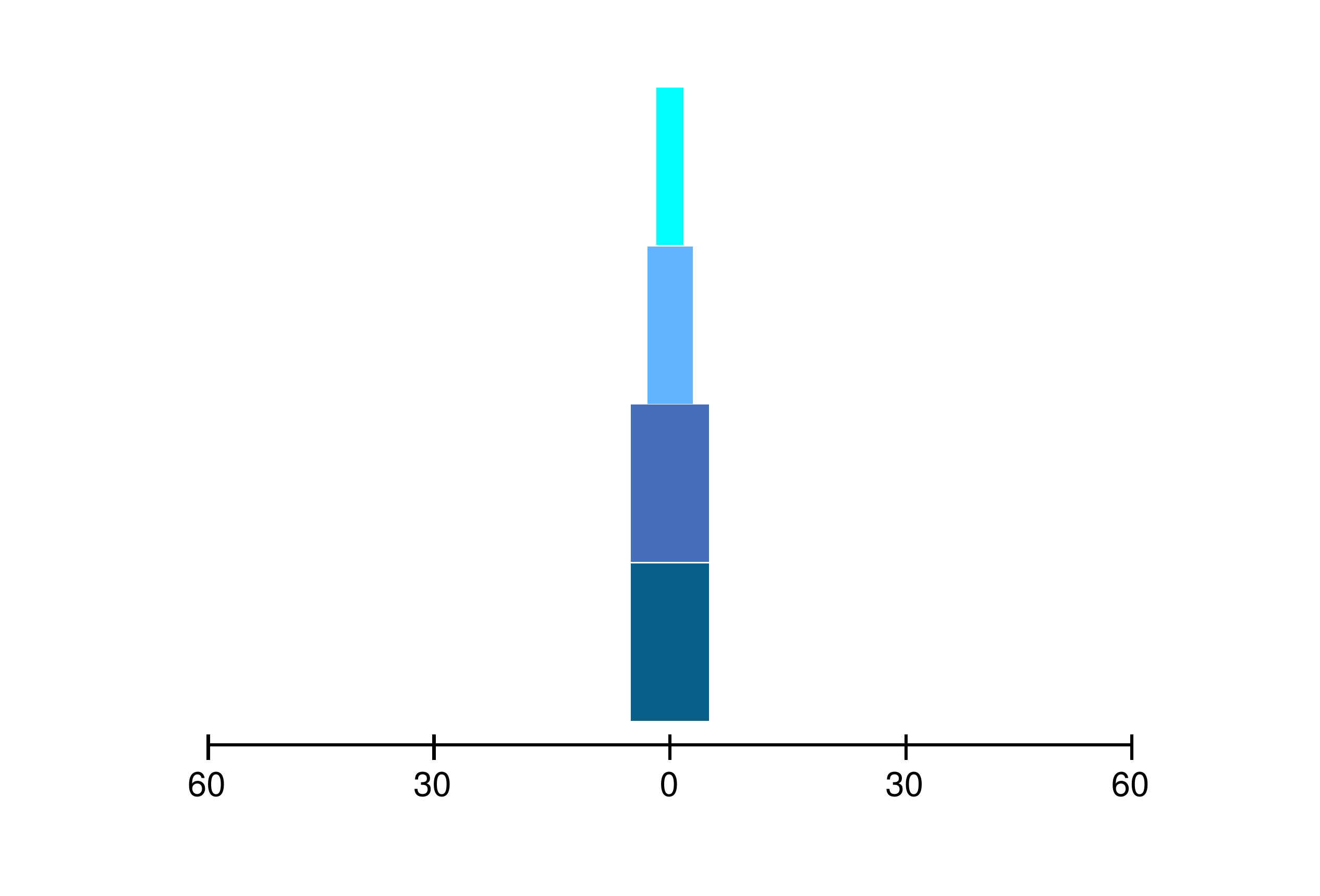}
\caption{Ecosystem profiles (mean number of species in each trophic level)
for increasing $\eta/\xi$ ratios. This plot corresponds to a resource saturation $R=1340$, 
for which communities have up to 4 trophic levels. Lower levels are shown with darker 
color. (a) The ecosystem maintain its pyramidal structure ($\eta/\xi=0.05$). (b) The 
first trophic level starts to collapse ($\eta/\xi=0.3$). (c) The second level starts to 
loose species ($\eta/\xi=0.43$). (d) For large values of the ratio, the system is close 
to extinction ($\eta/\xi=0.6$).}
\label{fig:profiles}
\end{center}
\end{figure}

A second signal of the transition in this model is a gradual loss of species in trophic 
levels from bottom to top. This effect can be qualitatively observed in the ecosystem 
profile (see \Fref{fig:profiles}), where the average number of species at each
trophic level is shown. When the quotient of rates increases from $\eta/\xi=0.3$
[panel (b)] to $\eta/\xi=0.43$ [panel (c)], the number of species in the first level
decreases considerably, but the rest of levels remain almost unaltered. After
that [panel (d)], a simultaneous loss of species in the first and second levels
takes place.

\begin{figure}[t!]
\begin{center}
\includegraphics[width=80mm,clip=true]{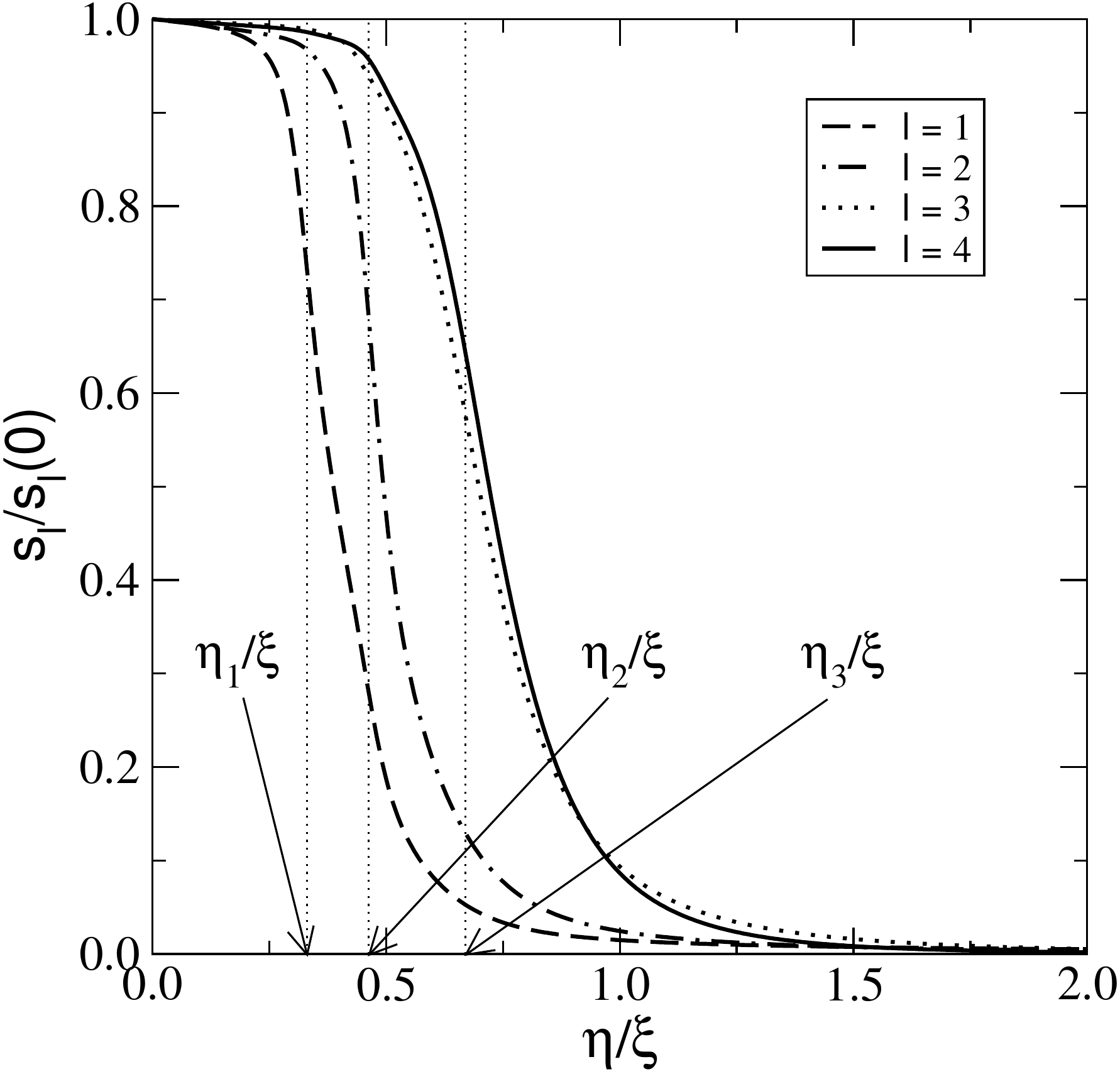}
\caption{Average number of species at each trophic level, $s_{\ell}=
\sum_{i\in\mathcal{G}}\pi_i s_{\ell}^{(i)}$, for $\ell=1,2,3,4$ and $R=1340$, normalized
by the average number of species for $\eta=0$, $s_{\ell}(0)$. We have marked the points 
$\eta_1$ and $\eta_2$ corresponding to the maxima of $\sigma_S^2$, which roughly coincide 
with the points at which the first and second levels start to collapse. The species loss 
in the third and fourth levels starts when the probability $P_{\varnothing}$ becomes 
appreciable (by $\eta_3$, the maximum of $\sigma_T^2/T^2$).}
\label{fig:speciesL}
\end{center}
\end{figure}

\Fref{fig:speciesL} shows the number of species at each level averaged over $\mathcal{G}$
versus $\eta/\xi$. We observe that the decrease of $s_1$ approximately coincides with
the first peak of $\sigma_S^2$ at $\eta_1$, and the decrease of $s_2$ corresponds to the
second peak at $\eta_2$. After that, species at lower levels are unable to sustain
upper levels and a trophic cascade occurs. The third and fourth levels start to be 
emptied near $\eta_3$. There is a clear correspondence between the values at which trophic 
levels start to collapse and the location of the maxima of $\sigma_S^2$ and 
$\sigma_T^2/T^2$. In any case, the loss of species from bottom to top as the extinction 
rate increases is a clear signal of the catastrophic regime shift. Besides, trophic 
cascades have been recognized empirically as signals of over-fishing in marine communities 
\cite{daskalov:2007}.

\begin{figure}[t!]
\begin{center}
\includegraphics[width=80mm,clip=true]{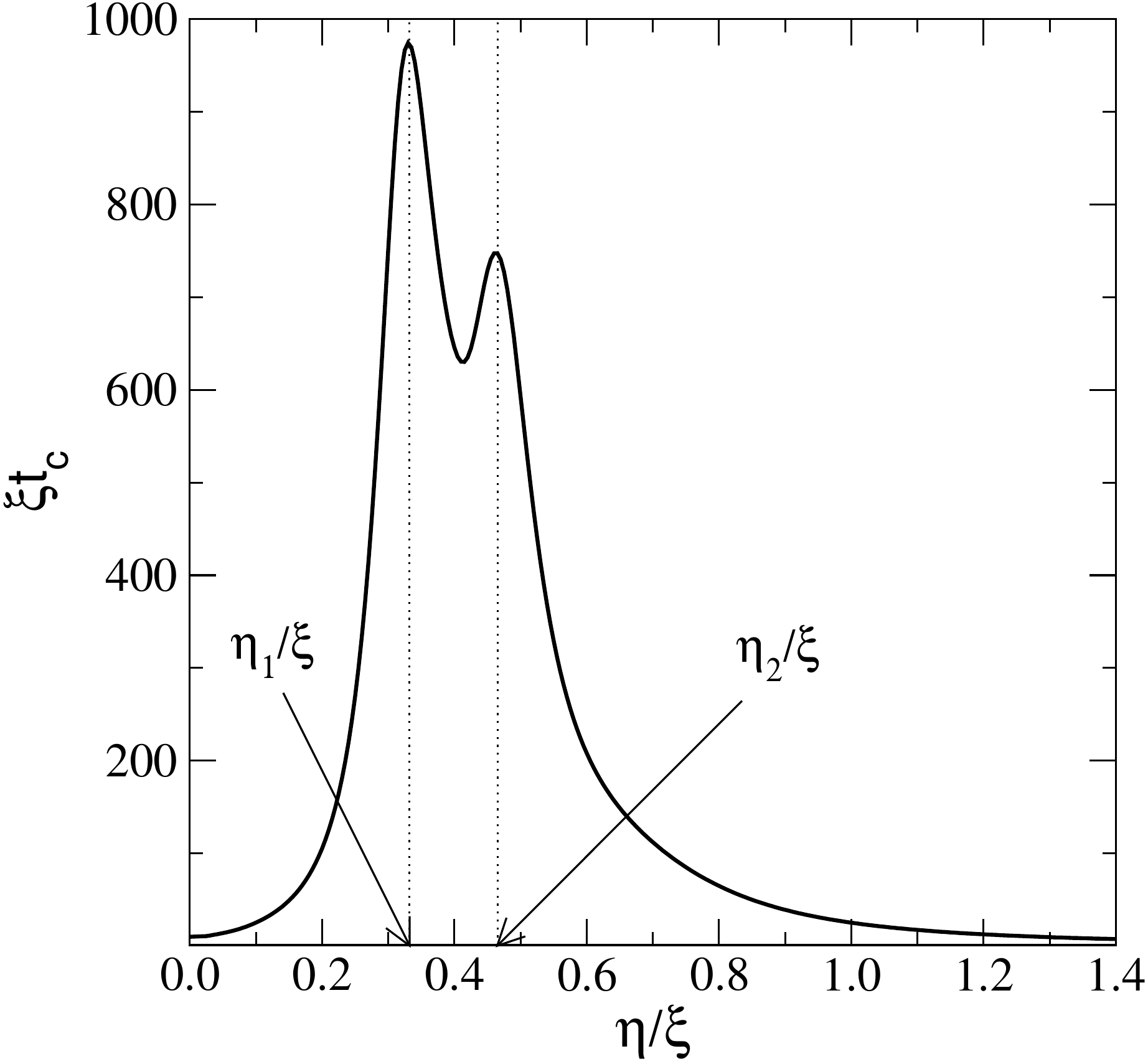}
\caption{Convergence time $\xi t_{\rm c} = -(\log|\lambda_2|)^{-1}$ needed to reach
the steady state of the Markov chain, $\lambda_2$ being the second largest eigenvalue 
of matrix $P$. Two peaks appear in $\xi t_{\rm c}$ corresponding to the two
maxima of $\sigma_S^2$.}
\label{fig:timestat}
\end{center}
\end{figure}

In the remaining of the section we will consider the variation of $\eta$ as a 
non-equilibrium process. Now we shall assume that, although the variation of
$\eta$ is still not faster than $\xi^{-1}$, the two scales are comparable in
the sense that the process is not able to remain in the steady state anymore.
An estimate of the time scale for the convergence to the steady state is 
provided by
\begin{equation}
\xi t_{\rm c} = -\frac{1}{\log |\lambda_2|},
\end{equation}
where $\lambda_2$ is the second largest eigenvalue (in modulus) of the
stochastic matrix $P$ (see \Fref{fig:timestat}). The distance from the
probability distribution after $n$ iterations of the Markov chain to the
steady state is proportional to $|\lambda_2/\lambda_1|^n$, hence the definition
of $t_{\rm c}$ (note that the maximum eigenvalue $\lambda_1=1$ since $P$ is a
stochastic matrix). For $R=1340$, the number of iterations needed to reach
equilibrium near the shift are around $10^3$.

\begin{figure}[t!]
\begin{center}
\includegraphics[width=80mm,clip=true]{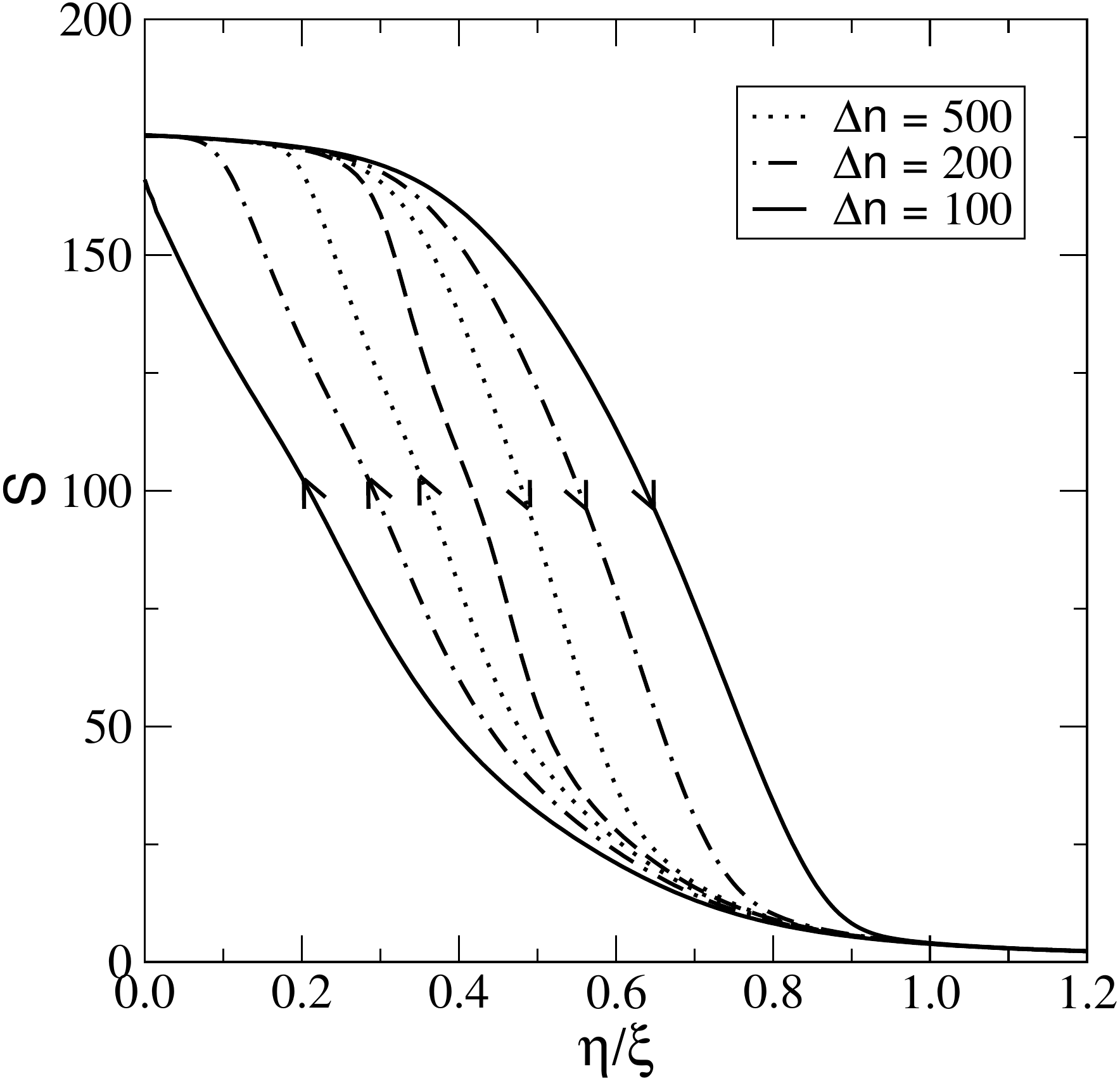}
\caption{Hysteresis cycles for $\Delta\eta/\xi=0.005$ and
three values of $\Delta n$ (see text). Dashed line represents the average
number of species in the quasi-stationary process of variation of $\eta$ (for
$\Delta n\rightarrow\infty$). The larger $\Delta n$ the further from
equilibrium the system and the narrower the cycle.}
\label{fig:hist}
\end{center}
\end{figure}

The faster variation of $\eta$ is implemented by producing a small change
$\Delta\eta$ every $\Delta n < \xi t_{\rm c}$ iterations of the Markov chain.
We start by increasing $\eta$ in these increments until reaching an arbitrary
value beyond the regime shift. Then we repeat the process by decreasing $\eta$
in the same increments. This way we can track any observable along the cycle,
by computing its averages after $k=0,1,\dots$ increments $\Delta\eta$ using the
probability distribution
\begin{equation}
\pi(k\Delta\eta/\xi)=\pi(0)P^{\Delta n}(0)P^{\Delta n}(\Delta\eta/\xi)\cdots
P^{\Delta n}(k\Delta\eta/\xi),
\end{equation}
given any initial distribution $\pi(0)$ at $\eta=0$ (reverse order in matrix 
products applies for decreasing $\eta$). In \Fref{fig:hist} the average
species richness exhibits a hysteresis cycle. As $\Delta n$ increases this
cycle narrows, recovering the quasi-stationary process in the limit
$\Delta n\rightarrow\infty$. In this limit the process is reversible and
the cycle collapses to the curve of mean number of species shown in
\Fref{fig:fluct}.

The existence of hysteresis loops in overexploited systems has been reported in the 
literature as another signal of catastrophic shifts in ecological communities 
\cite{scheffer:2001,fernandez:2009,scheffer:2009}. We have obtained it for the mean 
number of species, but a similar behavior will be observed in any other
magnitude (like the total biomass density, as in \cite{fernandez:2009}). 
In spite of the simplicity of our model, 
hysteresis cycles appear as well as other usual precursors of catastrophic regimes,
such as anomalous variance. But, unlike previous models, our model 
allows a deeper understanding of the transition, as we will discuss below.
An important insight this model provides is that hysteresis can arise as
a result of a time-scale mixing that keeps the system out of equilibrium
rather than to non-linearities of the underlying population model.

\section{Phase transition in finite size}\label{s:phase}

As we have mentioned, our model provides a full description of the phase space
of the system by means of a transition probability matrix. We are going to take 
advantage of this fact to show rigorously that the phenomenology that we have 
described in Section \ref{s:signals} corresponds to 
a phase transition ---in the sense of Statistical Mechanics--- in finite size. 

In Statistical Mechanics, phase transitions are associated to non-analyticities of the 
free energy of a physical system. Whenever the system is described by a transfer
matrix, the free energy is obtained from its largest eigenvalue.
A true phase transition would then be associated to the crossing of the
leading eigenvalue with the second largest (in modulus) one, because such 
a crossing causes a non-analytic behavior of the largest eigenvalue as a function
of the control parameter \cite{cuesta:2004}. The counterpart of a transfer 
matrix in a Markov chain is the transition probability matrix $P$. Thus a 
crossing of eigenvalues of $P$ would {\em rigorously} prove that the system
undergoes a phase transition.

Strictly speaking, the described shift can not be a true phase transition 
because of the finiteness of our system. According to the Perron-Frobenius theorem, an
irreducible matrix\footnote[1]{A matrix $P$ is reducible if there exists a permutation 
matrix $W$ such that $W^{\rm T}PW=\left(\begin{array}{cc}
X & 0\\
Y & Z
\end{array}\right).$ Hence the associated chain is not ergodic.}
with non-negative entries has a unique largest real eigenvalue 
and its corresponding eigenvector has strictly positive components \cite{meyer:2000}. 
Reducible matrices are related to processes with transient states. Since the 
Markov chain is ergodic for $\eta>0$, its stochastic matrix $P$ it is irreducible. 
Then the theorem implies that its maximum eigenvalue $\lambda_1=1$ is {\em simple}
for any value of $\eta/\xi$ (its corresponding {\em positive} left eigenvector is
precisely the asymptotic probability distribution). This excludes any eigenvalue
crossing, therefore any phase transition. True phase transitions can only
occur in infinite states Markov chains. When the limiting chain of a sequence
of finite Markov chains develops an eigenvalue crossing ---hence a phase 
transition--- the second largest eigenvalue of the elements of this sequence
approaches the largest one near the location of the true phase transition,
the more so the larger the size (number of states) of the Markov chain. This
is the fingerprint of the phase transition in finite systems. It is also
associated to a qualitative change in the eigenvector associated to the
leading eigenvalue.

\begin{figure}[t!]
\begin{center}
\includegraphics[width=80mm,clip=true]{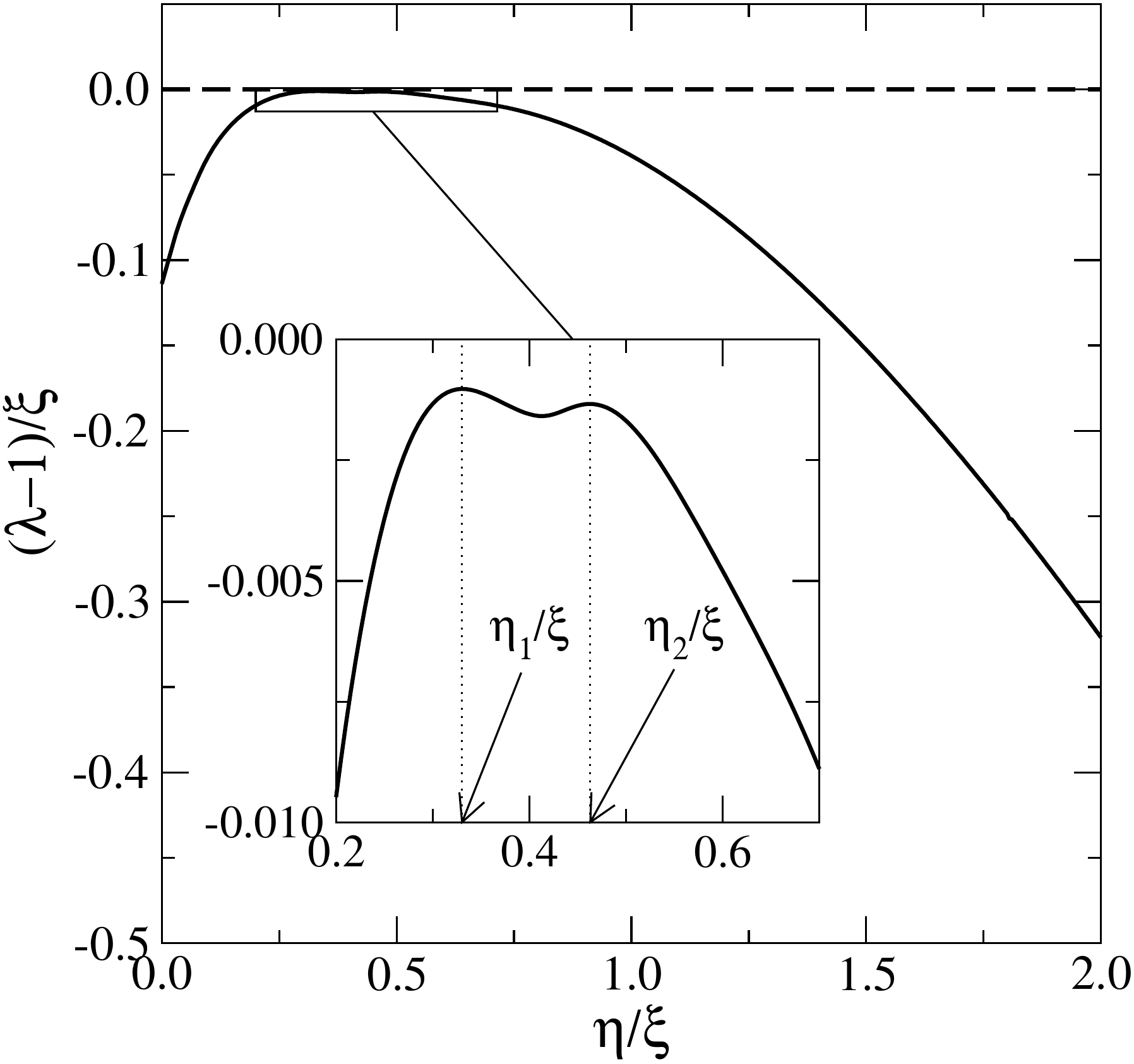}
\caption{First (dashed line) and second (full line) eigenvalues of $P$ for a resource
saturation $R=1340$. Inset shows a zoom of the region for which $\lambda_1=1$ and
$\lambda_2$
are closest, and two maxima appear at $\eta_1/\xi=0.330520\dots$ and $\eta_2/\xi=
0.462633\dots$. These points coincide with the maxima observed in $\sigma_S^2$ and 
$t_{\rm c}$ (see \Fref{fig:fluct} and \Fref{fig:timestat}).}
\label{fig:lambda2}
\end{center}
\end{figure}

We have computed the 10 largest eigenvalues of $P$ (in modulus) using Arnoldi iteration 
\cite{trefethen:1997} (useful for computing a few eigenvalues of large sparse 
matrices). In all cases the numerical method provides a real second
eigenvalue. \Fref{fig:lambda2} shows the dependence of $\lambda_2$ as a function of 
$\eta/\xi$ for $R=1340$. Not surprisingly, we identify the transition points as those 
of closest approach to the first eigenvalue (i.e., the two maxima observed in 
$\lambda_2$). Those maxima are reached at 
$\eta_1/\xi=0.330520\dots$ and $\eta_2/\xi=0.462633\dots$, which coincide with the 
values observed for the peaks in $\sigma_S^2$ and $t_{\rm c}$ (see Section
\ref{s:signals}). 
As we have shown before, each transition yields a trophic cascade
in the system, whose levels get emptied from bottom to top.

\begin{figure}[t!]
\begin{center}
\includegraphics[width=80mm,clip=true]{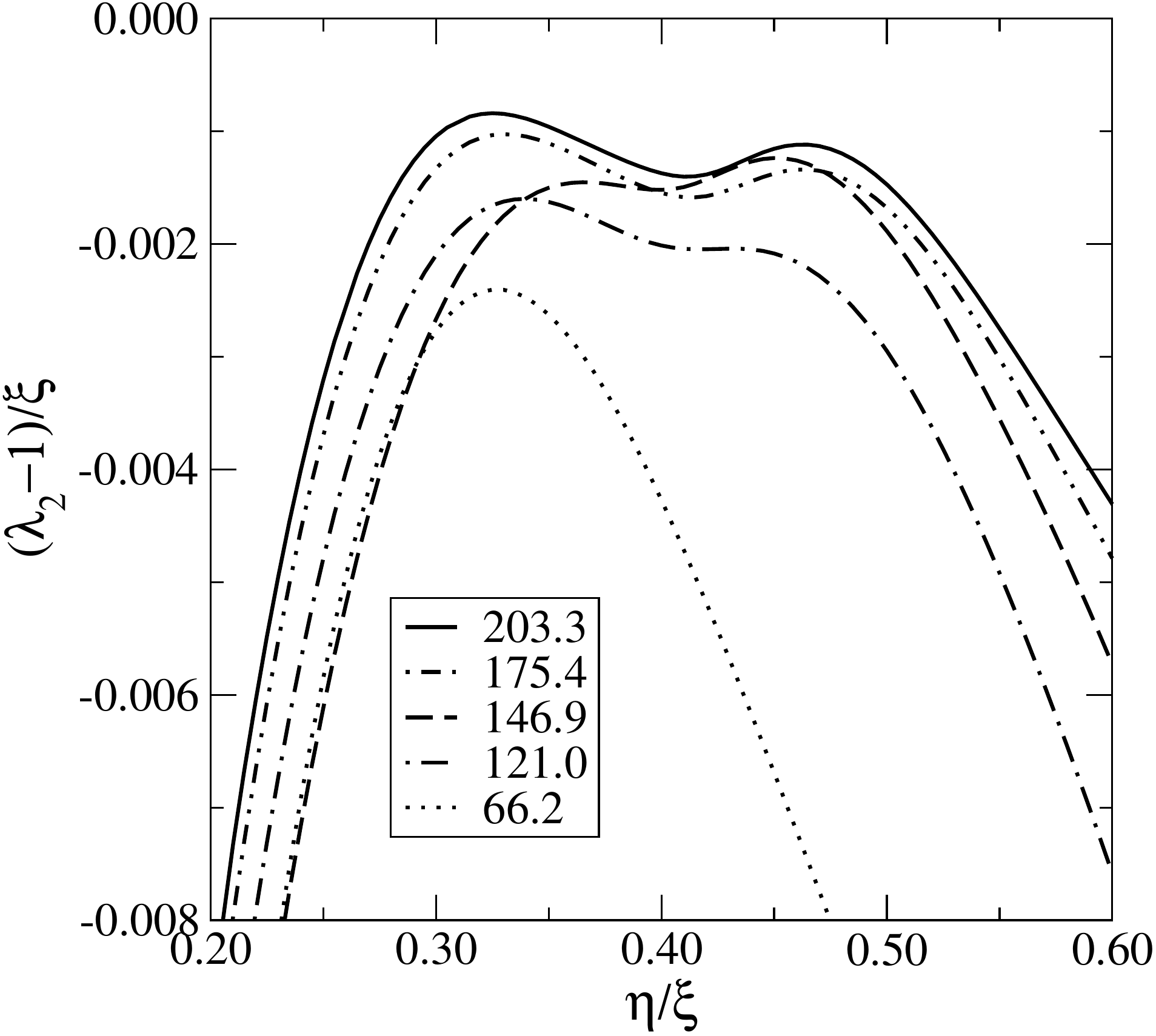}
\caption{Second eigenvalue of $P$ for several values of $R$
(legend shows the average number of species for each $R$ at $\eta=0$).
Dotted curve corresponds to $R=465$, for which the system only allows three
trophic levels and only a single maximum is observed. For higher values of $R$ (which
allow up to 4 levels), a second maximum appears, related to the 
trophic cascade in the second level. As the system size increases, $\lambda_2$
gets closer to 1.}
\label{fig:lambda2N}
\end{center}
\end{figure}

We have varied the system size (controlled by the amount of resource, $R$ 
\cite{capitan:2010b}) in order to check that the second eigenvalue gets closer to the 
first one as size increases. The system size is measured by the average number
of species in the recurrent set for $\eta=0$, and we do observe that the larger
the system size the closer is $\lambda_2$ to 1 (see \Fref{fig:lambda2N}). It
is difficult to increase the size beyond $S\approx 300$ because the number of
nodes in $\mathcal{G}$ grows approximately as $e^{\kappa\sqrt{R}}$ \cite{capitan:2010b}. 
For $S\approx 300$ the number of viable communities is larger than $10^6$, and the 
eigenvalue computation becomes very demanding.

\begin{figure}[t!]
\begin{center}
\includegraphics[width=80mm,clip=true]{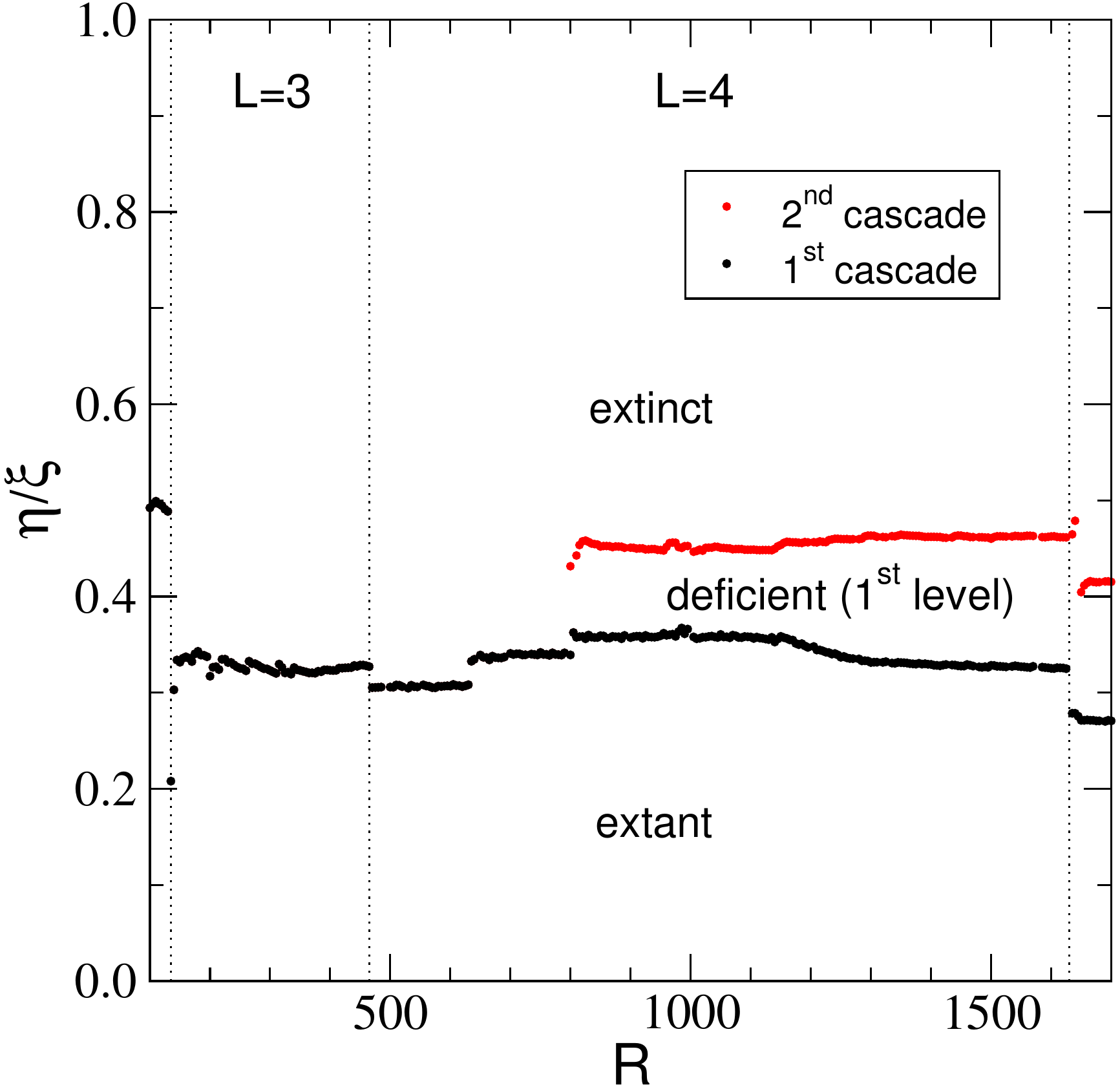}
\caption{Phase diagram of the system: black circles correspond to the
abscissa of the first maximum of $\lambda_2$ (trophic cascade at the first level), and 
red ones to the second maximum (trophic cascade at the second level). Above the second
transition, the process stays in communities close to the extinct ecosystem.}
\label{fig:phase}
\end{center}
\end{figure}

To conclude this section, we have obtained the phase diagram of this system.
In \Fref{fig:phase} we show in a $\eta/\xi$ vs.~$R$ diagram the
points at which $\lambda_2$ reaches its maxima, yielding the transition lines
corresponding to the different trophic cascades of the system. Basically transition 
lines are horizontal, and the number of undergone trophic cascades appears to be related 
to the number of trophic levels of the communities. We believe that, at some point in 
the region in which communities have 5 trophic levels,
a third maximum would appear related to a separate trophic 
cascade at the third level. Bearing this hypothesis out is difficult, though, 
because it would require eigenvalue calculations with too large matrices.

All this analytical evidence allows us to claim that these model ecosystems
undergo a catastrophic transition driven by the relative extinction rate. 
The dependence of $\lambda_2$ with $\eta/\xi$ actually exhibits two maxima
and so does the variance in species number. This points towards the existence
of a double phase transition, each one associated to a trophic cascade that 
collapses the lowest and next-to-lowest trophic level in the ecosystem.
Increasing the external stress over the 
system above these values increases the probability of driving the ecosystem 
to extinction. Consequently there is a threshold in $\eta/\xi$ above 
which the extinction of the ecosystem is the most likely event.
Small variations in that region can cause the collapse of stable ecosystems.

\section{Conclusions}\label{s:conclusions}

In this paper we have used a recently introduced model \cite{capitan:2009} to propose 
an alternative explanation to the observed catastrophic regime shifts in overexploited 
ecosystems \cite{scheffer:2001,fernandez:2009}. Previous models postulate 
nonlinearities in the 
dynamics of a magnitude representative of the whole ecosystem, like for instance the 
total biomass density, and this non-linear behavior leads to bi-stability. The system 
undergoes a transition due to a change of regime that leads the system from one stable 
state to another. This viewpoint is almost the same used in the classical explanations 
of the liquid-vapor transition in thermodynamics, for which the potential of the 
system has two alternative stable states.

Our model, however, is inspired in a microscopic description, more related to the 
perspective of Statistical Mechanics. Our viable communities are micro-states in 
a finite phase space and represent different states
of the ecosystem. The transition between a species-rich attractor (the recurrent set 
at $\eta=0$) and an attractor with low number of species (communities close to the
empty ecosystem) is 
explained as the crossing of two eigenvalues of the transition matrix of the Markov 
chain. This transition is driven by an increasing external force represented 
by a background species extinction rate, that takes into account the overall mortality
for reasons other than predation (overexploitation, habitat destruction,
epidemics\dots).

In spite of its being minimalistic, the
model reproduces qualitatively the phenomenology observed
in overexploited ecosystems. We have studied for this model the behavior 
of several early-warnings that announce the catastrophic shift, and we have
found the same behavior as that obtained both in previous models \cite{scheffer:2001,
fernandez:2009} and empirically \cite{scheffer:2009}. These features are
shared by many systems under the framework of the {\em elementary catastrophe theory}
\cite{thom:1975}.

Catastrophes have characteristic fingerprints. Some of the standard flags of 
catastrophic regimes are {\em modality}, {\em anomalous variance} and
{\em hysteresis} \cite{gilmore:1981}. These are precisely the signals we find in
our model. Our system is bimodal because it undergoes transitions between
an attractor of high species richness to another stable state with low species 
richness. Fluctuations in the mean number of species exhibit peaks at
the transition points, and take very large values compared to their values
far away from the transition. We thus have anomalous behavior of variances.
And the average number of species exhibits hysteresis cycles when the system
is kept out of equilibrium, as we have shown in Section \ref{s:signals}.
This explanation of hysteresis is alternative to the existence of non-linearities
in the population model, and points towards the speed of variation of the
external stress. The difference with the usual explanation is that in this
case the ecosystem can recover its initial state after releasing the external
stress, provided that we wait long enough and that there is availability of
invaders.

The main advantage of our model is the microscopic description that it provides.
The full characterization of the phase space of the system has allowed us
to show rigorously that the system undergoes a true phase transition
by computing the second eigenvalue of the transition matrix, which gets closest
to the first eigenvalue in the vicinity of the transition. This provides
a theoretical
support to the critical behavior exhibited by magnitudes like the fluctuation
of species richness.

We have found evidence for a double phase transition in the system,
associated to the gradual loss of species from bottom to top. This effect
is new to former models and could be used as early warning for the catastrophic 
shift by monitoring the species abundance at low trophic levels in overexploited
ecosystems. Trophic cascades have been revealed as possible mechanisms of 
catastrophic shifts in natural communities, though \cite{daskalov:2007}. 
Thus, with the caveat that this is just a very simplified picture
of real communities, our analysis of this model predicts
that overexploited systems will begin to collapse first at lower levels.

Assuming a different extinction rate at each trophic level would
be more realistic. For example, in overexploited marine ecosystems, the impact
of fishing pressure is stronger in higher trophic levels. A simple model like
ours could shed light in determining whether a strong extinction pressure in 
higher levels is more harmful than in lower levels. Refined versions of the
model could allow investigating this kind of effects.

Most theoretical explanations of the catastrophic phenomena observed in 
ecological systems subject to high exploitation pressure rely on nonlinearities in 
the macroscopic dynamics of the system. The take-home message of this paper is that 
a microscopic model like ours, based in ecologically reasonable assumptions
and in a simple, linear population dynamic models, can exhibit the same phenomenology
as non-linear, macroscopic models. This way, our approach 
can serve as an alternative explanation of the catastrophic regime shifts observed 
in ecological communities.

\section*{Acknowledgements}

This work is funded by projects MOSAICO, from Ministerio
de Educaci\'on y Ciencia (Spain) and MODELICO-CM, from Comunidad Aut\'onoma 
de Madrid (Spain). The first author also acknowledges financial support through a contract
from Consejer\'{\i}a de Educaci\'on of Comunidad de Madrid and Fondo Social 
Europeo.

\appendix
\section*{Appendix}
\setcounter{section}{1}

This appendix is devoted to calculate the fluctuation of the average
time of first return to the empty ecosystem. Throughout this section
$f_{ij}^{(n)}$ will stand for the probability that in a process starting form the state
$i$ the first entry to $j$ occurs after $n$ steps. The distribution $\{f^{(n)}_{ij}\}$
is known as the first-passage distribution for the state $j$ (in particular, 
$\{f^{(n)}_{jj}\}$ represents the distribution of the recurrence times for $j$).
This distribution is related to the probabilities $p_{ij}^{(n)}$ of a transition
from $i$ to $j$ in exactly $n$ steps \cite{feller:1968} according to
\begin{equation}\label{eq:pf}
p_{ij}^{(n)}=\sum_{\nu=1}^n f_{ij}^{(\nu)}p_{jj}^{(n-\nu)}.
\end{equation}
Let us study the case $i=j=\varnothing$. We are interested in calculating the variance
of the recurrence time for the empty community, whose mean value is given by
$T=\sum_{n=1}^{\infty}nf_{\varnothing\varnothing}^{(n)}$.
To this purpose we introduce the generating functions $V(z)= \sum_{n=0}^{\infty} 
p_{\varnothing\varnothing}^{(n)}z^n$ ($p_{\varnothing\varnothing}^{(0)}=1$)
and $F(z)=\sum_{n=0}^{\infty} f_{\varnothing\varnothing}^{(n)}z^n$ 
($f_{\varnothing\varnothing}^{(0)}=0$). Then \eref{eq:pf} is equivalent 
\cite{feller:1968} to
\begin{equation}\label{eq:FU}
F(z)=1-\frac{1}{V(z)}
\end{equation}
and $T=F'(1)$. It can be easily shown that the radius of convergence of $V(z)$
is equal to 1, and that $V(1)$ diverges by definition. By imposing that $F'(1)$
is finite we find that, near $z=1$, $V(z)\approx a(1-z)^{-1}$ and 
$a=\lim_{n\rightarrow\infty}p_{\varnothing\varnothing}^{(n)}=P_{\varnothing}$. Hence the 
mean recurrence time for $\varnothing$ is $T=F'(1)=a^{-1}=P_{\varnothing}^{-1}$.

The variance of the recurrence time is obtained as
\begin{equation}
\sigma_T^2=\sum_{n=0}^{\infty}n^2f_{\varnothing\varnothing}^{(n)}-T^2
=F''(1)-F'(1)[1-F'(1)].
\end{equation}
In order to calculate $F''(1)$, we need to obtain the next-to-leading (constant)
term in the series expansion of $V(z)$ in powers of $1-z$,
$V(z) = P_{\varnothing}(1-z)^{-1}+b+ \mathcal{O}(1-z)$. Using \eref{eq:FU} we
get $F''(1)=2b/P_{\varnothing}^2$ and
\begin{equation}
\frac{\sigma_T^2}{T^2}=2b+P_{\varnothing}-1.
\end{equation}
To conclude with this calculation, we need to obtain a way to compute
numerically the constant $b$. This is an easy task, because
\begin{equation}
b+\mathcal{O}(1-z)=V(z)-\frac{P_{\varnothing}}{1-z}=\sum_{n=0}^{\infty}
\left(p_{\varnothing\varnothing}^{(n)} - P_{\varnothing}\right)z^n.
\end{equation}
Hence, in the limit $z\rightarrow 1$,
\begin{equation}
b=\sum_{n=0}^{\infty}\left(p_{\varnothing\varnothing}^{(n)}-P_{\varnothing}\right).
\end{equation}
Therefore we simply need to iterate the matrix $P$ and truncate the series 
up to certain error tolerance to compute $b$. 

\section*{References}

\bibliographystyle{unsrt}
\bibliography{ecology}

\end{document}